\documentclass[a4paper,fleqn,usenatbib]{mnras}

\usepackage[T1]{fontenc}
\usepackage{ae,aecompl}
\usepackage{pdflscape}

\usepackage{graphicx}	
\usepackage{amsmath}	
\usepackage{amssymb}	

\title[Statistical analysis of properties of dwarf novae outbursts]{Statistical analysis of properties of dwarf novae outbursts}

\author[M. Otulakowska-Hypka, A. Olech, and J. Patterson]{
Magdalena Otulakowska-Hypka,$^{1}$\thanks{E-mail: magdaot@strw.leidenuniv.nl}
Arkadiusz Olech,$^{2}$
and Joseph Patterson$^{3}$
\\
$^{1}$Leiden Observatory, Leiden University, P.O. Box 9513, NL-2300 RA Leiden, The Netherlands\\
$^{2}$N. Copernicus Astronomical Center of the Polish Academy of Sciences, ul. Bartycka 18, 00-716 Warsaw, Poland\\
$^{3}$Department of Astronomy, Columbia University, 550 West 120th Street, New York, NY 10027, USA
}

\date{Accepted XXX. Received YYY; in original form ZZZ}

\pubyear{2016}

\begin{document}
\label{firstpage}
\pagerange{\pageref{firstpage}--\pageref{lastpage}}
\maketitle

\begin{abstract}
We present a statistical study of all measurable photometric features of a large sample of dwarf novae during their outbursts and superoutbursts. We used all accessible photometric data for all our objects to make the study as complete and up-to-date as possible. Our aim was to check correlations between these photometric features in order to constrain theoretical models which try to explain the nature of dwarf novae outbursts. 
We managed to confirm a few of the known correlations, 
that is the Stolz and Schoembs Relation, 
the Bailey Relation for long outbursts above the period gap, 
the relations between the cycle and supercycle lengths,
amplitudes of normal and superoutbursts,
amplitude and duration of superoutbursts, 
outburst duration and orbital period,
outburst duration and mass ratio for short and normal outbursts, as well as 
the relation between the rise and decline rates of superoutbursts.
However, we question the existence of the Kukarkin-Parenago Relation but we found an analogous relation for superoutbursts.
We also failed to find one presumed relation between outburst duration and mass ratio for superoutbursts. 
This study should help to direct theoretical work dedicated to dwarf novae. 
\end{abstract}




\section{Introduction}
Dwarf novae (DN) are one of the types of cataclysmic variables (CVs), composed of a white dwarf accreting material from its close companion -- a main-sequence star. 
Based on their photometric behavior, we distinguish a few sub-classes of DN: 
U~Gem (showing more or less similar outbursts), 
SU~UMa (characterized by the so-called superoutbursts next to normal outbursts), and 
Z~Cam stars (with outbursts interrupted by irregular standstills, i.e., intervals of constant brightness). 
Within the SU~UMa class there is an additional distinction -- from the most to the least active ones: ER~UMa-type, pure SU~UMa-type, and WZ~Sge-type stars. See \cite{2001Hellier} and \cite{2003Warner} for a detailed overview. 

There exists a number of theoretical models which attempt to explain, among other things, the nature of outbursts, superoutburst and superhumps of dwarf novae. The most popular are the following ones.

The \textit{thermal limit-cycle instability model} \citep{1981Meyer, 1984Meyer, 1984Smak, 1993Cannizzo, 2001Lasota, 2010Cannizzo} showing that all outbursts of SU~UMa-type stars and of all other types of DN have essentially the same nature. They are understood in terms of 
the pure thermal limit-cycle instability in their accretion disks, without tidal effects.
This principle is usually presented in the form of the \textit{S-curve plot} (the surface density, $\Sigma$, vs. surface temperature, $T$, or viscosity or mass transfer rate). 
Both horizontal branches of the S-curve are stable: the lower one represents the cold and low-viscosity disk of a quiescent DN, whereas the upper branch shows the hot and high-viscosity disk in an outburst. 
According to this model, a DN outburst is caused by a propagation of a heating wave in the accretion disk rather than an outburst of the whole disk at once. 
All eruptions of all DN are understood as thermal oscillations between both stable states on the S-curve.
In this model, normal outbursts and superoutbursts of SU~UMa-type stars are thought to be analogous to "narrow" and "wide" outbursts of longer orbital period DN. 
According to the classification of \cite{1984Smak}, normal outbursts are of the Type Bb, i.e., the disk's outer edge does not expand and only the mass in the inner part is accreted, while the mass in the outer part of the disk is untapped. 
In turn superoutburst are interpreted as Type Ba outburst when the heating wave reaches the disk's outer edge.
Shapes of all outbursts, their duration and rates of rise depend on the radius at which the outburst is triggered, the type of an outburst (inside-out or outside-in), and on the distribution of material left by the previous outburst. 
The declines of all outbursts are alike, since the cooling wave always moves inwards in the disk \citep{2001Hellier}. 

Another proposed model is the so-called \textit{thermal-tidal instability (TTI) model} \citep{Osaki1974, 1988Whitehurst, 1996Osaki}  which, apart from the thermal instabilities, takes into account also tidal instabilities in accretion disks of DN. 
The main difference between this and the previous model is that normal outbursts are understood here as caused by the thermal instability alone, while during superoutburst, the tidal instability additionally occurs. 
TTI model interprets the length of supercycles ($P_{sc}$, i.e., time between two consecutive superoutbursts) as inversely proportional to the mass transfer rate ($\dot{M}_{tr}$). 
The material in the disk is not completely removed due to normal outbursts because more material has been accreted within such a cycle and, as a result, matter continues to accumulate during a number of normal outbursts. 
After such a disk reaches the critical radius ($r \sim 0.46 a$, where $a$ is the binary separation) it becomes eccentric and then the tidal effects play an important role. It is the main cause of superoutbursts, after which the disk shrinks again. 
The tidal instability is caused by the 3:1 resonance between the flow of material in the disk and the orbital motion of the secondary component. 
An important limitation of this model is that the critical resonance radius occur only for systems with the mass ratio $q=M_2/M_1$ smaller than 0.25, i.e., only for CVs with $P_{orb}\leqslant2$~h. 

Last but not least, there is \textit{the enhanced mass-transfer (EMT) model} \citep{1983Vogt, 1985Osaki, Smak2004c, Smak2004a, 2008Smak}. 
This model is based on the observations of a higher mass transfer rate just before a superoutburst, as \cite{1983Vogt} observed as increased amplitude of orbital humps during normal outbursts, and the fact that the accretion rate, the irradiation from the secondary, and the mass outflow rate are strongly correlated. 
In this model, superoutbursts of SU~UMa stars are explained by enhanced mass transfer from the secondary star. The EMT is caused by irradiation heating of the atmosphere of the secondary star due to ultraviolet radiation from the mass-accreting WD and the boundary layer between the accretion disk and the WD \citep{Smak2004c}.
Superoutbursts' appearance does not depend on the outer radius reached during the expansion of the disk, as in the TTI model, but depends on the maximum accretion rate during outburst which increases with the mass of the disk. Thus the irradiating flux and the maximum mass transfer rate also depend on how much mass has been stored in the disk \citep{2004Schreiber}.

Although the TTI model is commonly accepted, it still seems to need some improvements. For instance, U~Gem and Z~Cam stars also show superoutbursts, although very rarely \citep{2001Cannizzo, 2002Cannizzo, 2004SmakSmak}.
It seems that during such long outbursts they accrete more material than was stored in the disk, which can be explained by an enhanced mass transfer.
There exist also cases of permanent superhumpers above the period gap as well as nova-like objects below the gap \citep{2005Osaki, 2012Patterson}.
Another example is the fact shown by \citet[and references therein]{2012Cannizzo} that embedded precursors of superoutbursts are present also in systems with long orbital periods. This argues for a more general model, which is not restricted to the mass ratio of $q<0.25$.
What is more, TTI simulations which reproduced the light curve of one of the most active dwarf novae, ER~UMa star, required an artificial increase of the mass transfer rate by the factor of ten in comparison to values expected from the theory based on gravitational radiation \citep{1995Osaki}.
Recently \citet{2012Osaki} presented evidence in favor of the TTI model.
Although their analysis was based on variations of the negative superhump period of a single dwarf nova, V1504~Cyg, they claimed that the cause of superoutbursts was finally revealed  to be
pure TTI.
However, \citet{2013Smak} showed that this object cannot be considered as a representative for all systems which show negative superhumps. He also presented a number of arguments which suggest that superoutbursts are caused by strongly enhanced mass transfer rate \citep{1991Smak, 2008Smak}. 

Outbursts of DN do not strictly reproduce, i.e. in a given object there are no two identical outbursts. In spite of this fact, there are some systematic effects present in their photometric characteristics and we aim to analyze them in this work as well as all possible correlations between these measurable quantities. 
In the literature there is a number of suggestions for the existence of such correlations. Some of them are confirmed, for example the Kukarkin-Parenago relation between the amplitude of a normal outburst and the period between outbursts \citep{1934Kukarkin}, and the Bailey relation between the rate of the brightness decrease in an outburst and the orbital period \citep{1975Bailey}. Also, some of these relations are not confirmed, such as the suggestion in the work of \cite{2007AcA....57..267R} that stars with longer orbital periods less likely show the so-called secondary humps in their superhumps. 

Despite the already rich observational material for DN, there is still no up-to-date data base gathering \textit{all} the measurable photometric features of these stars, and their analyses. 
Recently, the first such attempt was presented by \cite{2015arXiv151203821C} where the CRTS data was used to determine the apparent outburst and quiescent V-band magnitudes, duty cycles, limits on the recurrence time, upper- and lower-limits on the distance and absolute quiescent magnitudes, colour information, orbital parameters, and X-ray counterparts.
Many authors focus mainly on the periodicities of dwarf novae, but overlook other basic information which can be obtained from the light curves, such as their amplitudes. For some of these objects we know values of their basic parameters, such as mass ratio and orbital period, but there is still no comprehensive and systematic analysis of the whole sample. This was our motivation for this work. We created such a data base which is now public\footnote{\texttt{http://users.camk.edu.pl/magdaot/DN.html}} and performed its first statistical analysis which we present here. 

In the next Section we present the data sources used, together with a description of our measurements and results in the form of a catalog. 
Section~\ref{sec-relations} shows our attempt to test possible relations between the measured parameters. 
In the last Section~\ref{sec-relations-con} we conclude the study of our sample of dwarf novae.

\section{Data}
\label{sec-data}

\subsection{Data sources}
\label{data-sources}

In this study we used the following available catalog data sources: 

\begin{enumerate}
\item \textit{The Catalog and Atlas of Cataclysmic Variables}\footnote{\texttt{https://archive.stsci.edu/prepds/cvcat/}} by \cite{2001Downes}
which contains 1830 objects that have been classified as a CV before Feb.~1, 2006, when the catalog was frozen. 

\item \textit{Catalogue of cataclysmic binaries, low-mass X-ray binaries and related objects}\footnote{\texttt{http://www.mpa-garching.mpg.de/RKcat/}}
 by \cite{2003Ritter}. Although the reference corresponds to a catalog which is over 10 years old, its newest edition 7.21 (Dec.~31, 2013) have been used in this study. This catalog contains 1094 CVs. 

\item  Catalog of J.~Patterson, that is the
Supplementary Electronic Material to the publication \cite{2011Patterson} containing properties of 292 non-magnetic CVs with orbital periods smaller than 3~hours\footnote{\texttt{http://cbastro.org/dwarfnovashort/}}.

\end{enumerate}

The catalog of \cite{2001Downes} provides the following parameters: 
object name, coordinates, proper motion, galactic coordinates, 
the year of outburst (for novae), the maximum and minimum magnitudes for non-novae objects from the catalog of Ritter and Kolb, references, 
orbital period, space-based observations, a flag indicating if the object is in a globular or open cluster,  finding charts, and the type of variability.

In turn, the catalog of \cite{2003Ritter} contains: 
object name, coordinates, 
orbital period, superhump period, flag indicating the occurrence of eclipses, type of spectroscopic binary, spectral types, eccentricity of the orbit, mass ratio M1/M2, orbital inclination, masses and radii of both components,
type and subtype of CV, 
apparent V magnitudes of:
Mag1 (DN stars in minimum), Mag2 (brightness at mideclipse, in case of eclipses), Mag3 (U~Gem and Z~Cam in outburst and SU~UMas in normal outburst), Mag4 (max. brightness of superoutburst or in standstill for Z~Cam stars), 
and cycles and supercycles lengths:
T1 (typical time interval between two subsequent outbursts for U~Gem and Z~Cam stars, or two subsequent normal outbursts for SU~UMa stars), and T2 (typical time interval between two subsequent superoutbursts for SU UMa stars).

To deal with such a large data set, the TOPCAT tool developed by \cite{TOPCAT} was used. 
TOPCAT\footnote{\texttt{http://www.star.bris.ac.uk/$\sim$mbt/topcat/}} stands for \textit{Tool for OPerations on Catalogues And Tables}. It is a Java-based interactive graphical viewer and editor for tabular data. 
It turned out to be the most convenient and powerful software for our purposes.  

Both catalogs of \cite{2001Downes} and \cite{2003Ritter} were 
downloaded from the VizieR catalog access tool\footnote{\texttt{http://vizier.u-strasbg.fr/}} into TOPCAT. 
They were matched within TOPCAT via right ascension ($\alpha$) and declination ($\delta$) with the algorithm "sky". 
The match selection criterion was set to the "symmetric best match" which 
treats the two tables symmetrically. Thus, any input row which appears in one result pair is disqualified from appearing in any other result pair. Every row from both input tables appeared in at most one row in the result table. 
In total 610 dwarf novae were found out of the matched pairs.

Sometimes the determination of the type of CV is not obvious, so there appeared some discrepancies in the classification of objects in both catalogs. The final types were taken from the last version of the catalog of \cite{2003Ritter}, since it is the most recent one.  What is more, not all dwarf novae have got a further classification into subtypes, so it is only given when available. 

The resulting matched catalog of 610 dwarf novae was expanded with the supplementary data for 292 CVs from \cite{2011Patterson}, which includes:
object name, max-min range of the V magnitude, orbital and superhump periods, Galactic coordinates, 
variable-star type, superoutburst recurrence period, inclination, mass ratio M2/M1, equivalent width of H$\beta$ emission line, proper motion, $\gamma$-velocity, estimated distance, estimated absorption, 
duration and V magnitude of the \textit{square-wave} equivalent of the plateau segment of
an average superoutburst, FUV and NUV magnitudes, and $M_V$. 
The missing parameters in our catalog such as magnitude levels, periods, mass ratio, inclination or supercycles lengths were updated when available.  
Particularly interesting and unique information for the purposes of this study are approximations of superoutbursts shapes with the \textit{square wave}, thus they were added to our catalog.

It is worth mentioning that of these various catalogues, some contain strictly observable quantities (\cite{2001Downes} is pretty close to that), others are a mix of observable and deduced quantities \citep{2011Patterson}, and others are mainly a list of published numbers \citep{2003Ritter}.
This heterogeneity probably plays a role in increasing the variance
found in our results. 

The catalog was further expanded by the analysis of the following photometric data. 
\begin{enumerate}
\item  Light curves from the database with amateur observations of \textit{the American Association of Variable Star Observers}\footnote{\texttt{http://www.aavso.org/}} for 357 objects. 
\item 43 dwarf novae of the OGLE\footnote{\texttt{http://ogle.astrouw.edu.pl/}} team:
three of them located in the Galactic bulge \citep{2011Poleski} and 
forty more in the OGLE-III Galactic disk fields \citep{2013Mroz}. \\
In turn the most recent photometry on over one thousand new dwarf novae from the OGLE Survey \citep{2016arXiv160102617M} was not used in this research, since it was published when our analysis was already completed. We plan to take this new data into account in our next, supplementary, analysis.
\item 
Light curves from the NASA's Kepler satellite\footnote{\texttt{http://archive.stsci.edu/kepler/}} for six objects presented by \cite{2012CannizzoCannizzo, 2012Ramsay} and \cite{2013Kato}. 
\item
Light curves from the database with amateur observations of 
\textit{the Center for Backyard Astrophysics}\footnote{\texttt{http://cbastro.org}} \citep{2003Patterson, 2005Patterson} for four objects. 
\item
Additional photometric data from our observational campaigns for eight objects (\cite{2003AcA....53..175O}, \cite{2004AcA....54..233O}, \cite{2006A&A...452..933O}, \cite{2008AcA....58..131O}, \cite{2009A&A...497..437R}, \cite{2009MNRAS.399..465O}, \cite{2011A&A...532A..64O}, \cite{2013MNRAS.429..868O}). 
\end{enumerate}

This photometric data provides light curves mostly in the V-band. The only exceptions are OGLE data in the I-band and KEPLER light curves in flux units. KEPLER data was transformed into magnitudes according to the description given in \cite{2013MOH-Psc}.

We collected a big sample of stars for which we have photometry of miscellaneous quality, in best cases it is covering tens or hundreds of days of almost continuous observations. The data was subject to a careful and systematic analysis, presented in the next subsection.

\subsection{Photometric features of our interest}
\label{phot-feat}

Extensive analysis was performed for the sample of over 400 light curves from all sources listed in the previous subsection, that is AAVSO, OGLE, Kepler, CBA, and our data.
For each of them every measurable parameter from the list given below was recorded into our catalog. 
Each of these light curves has been investigated manually because of several reasons. 
First, there is no efficient automatic tool to perform such an investigation where all the features of outbursts could be recorded. 
Second, even if there was a fitting algorithm to do this task, it would work only for light curves of a particular quality, since such a numerical solution depends on the density of data points which is highly variable in our sample. Such an automatic analysis would require a manual examination afterwards in any case.
Third, the manual method of light-curve analysis is widely used in our community.

There is a \textit{bimodality} in duration of outbursts of DN. It is most prominent for the SU~UMa-type stars with normal and superoutbursts, but it is also present for many U~Gem and Z~Cam stars (with so-called short and long outbursts), however, it may not be clearly distinguishable on short light curves. 
None of the known catalogs presented in the previous section gives such a distinction of outbursts' duration and amplitude for U~Gem and Z~Cam stars. 
Nevertheless, in our analysis we separately recorded the characteristics of both short and long outbursts for these two classes, when it was possible. 
Our motivation to do this was the fact that we did not want to favor any of the existing theories explaining the nature of DN outbursts. Omitting this discrimination of short and long outbursts for U~Gem and Z~Cam stars would favor the TTI and EMT theories at pure thermal instability model's cost. 

When it comes to the outbursts' frequencies, they are not strictly periodic \citep{2012CannizzoCannizzo, 2013MOH-Psc} and for a given object the recurrence time can change over a factor of 2-3 (e.g. for V344~Lyr).
However, over many decades of observations the recurrence times are characteristic for each dwarf nova, so cycle and supercycle lengths based on long light curves give fairly stable statistics \citep{2003Warner}. 

\begin{figure}
\centering
\includegraphics[width=0.45\textwidth]{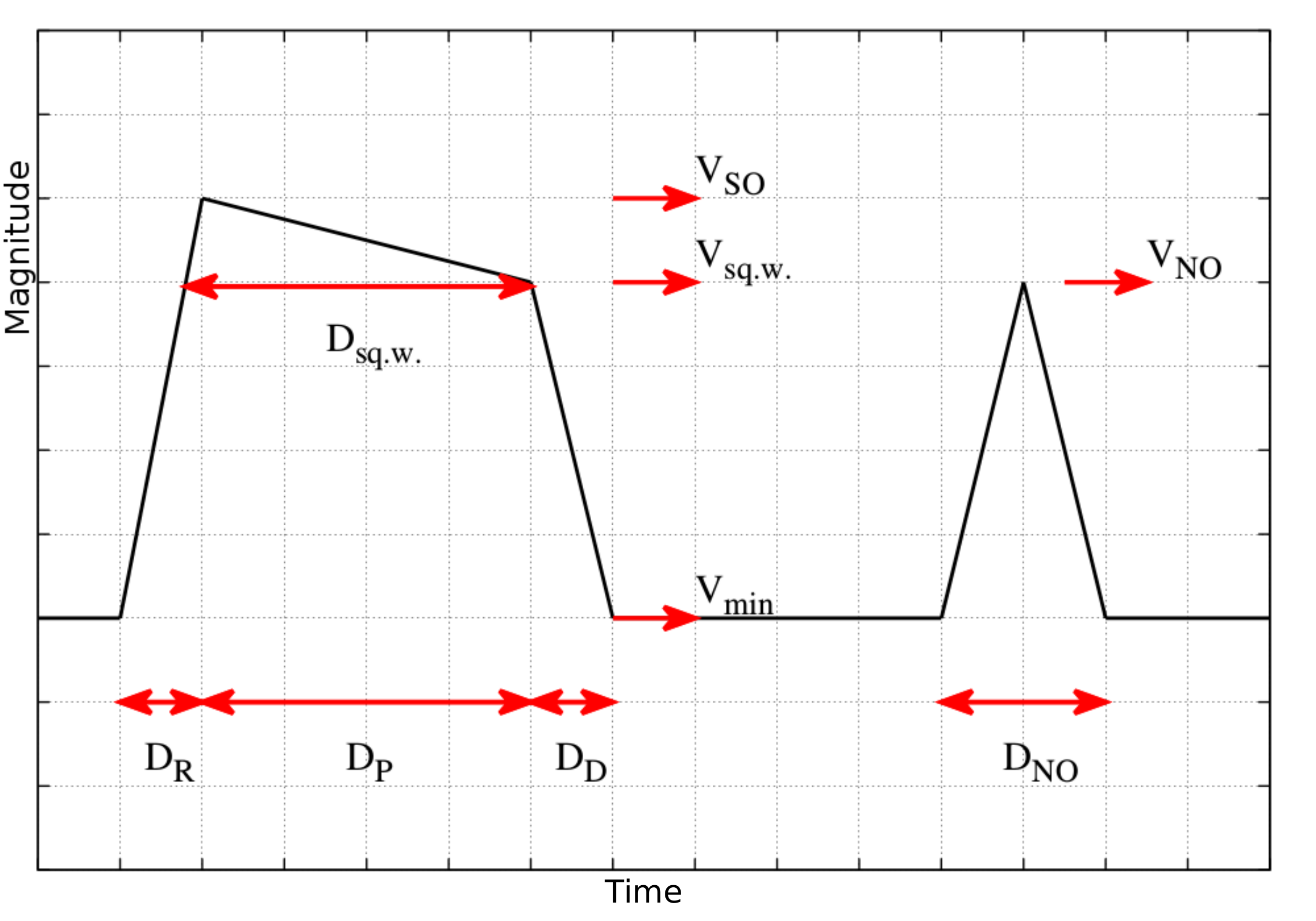}
\caption{Schematic shape of a light curve showing a superoutburst (in the case of SU~UMa stars) or a long outburst (in the case of U~Gem stars) and a following normal outburst with marked brightness levels and durations of particular phases which were measured.}
\label{fig-shape}
\end{figure}

While examining the light curves, the following characteristics were recorded and updated or added to our catalog. They are shown in Fig.~\ref{fig-shape}.

\begin{enumerate}
\item Apparent V magnitudes of normal outbursts and superoutbursts:
\begin{itemize}
\item $V_{min}$ -- maximum brightness of DN stars in minimum, 
\item $V_{SO}$ -- maximum brightness of superoutburst of SU~UMa stars or maximum brightness of long outbursts of U~Gem or standstill of Z~Cam stars (to maintain consistency with \citealt{2003Ritter}), 
\item $V_{NO}$ -- maximum brightness of normal outburst of SU~UMa stars or maximum brightness of short outbursts of U~Gem and Z~Cam stars.
\end{itemize}

\item Duration of consecutive three phases of super- or long outbursts:
\begin{itemize}
\item $D_R$ -- duration of a rise of an outburst, 
\item $D_P$ -- duration of the plateau phase, 
\item $D_D$ -- duration of an outburst's decline.
\end{itemize}

\item Duration of outbursts:
\begin{itemize}
\item $D_{NO}$ -- duration of a normal outburst of SU~UMa stars or a short outburst of U~Gem and Z~Cam stars,
\item $D_{SO}$ -- duration of a superoutburst of SU~UMa stars or a long outburst of U~Gem and Z~Cam stars, defined as $D_{SO}=D_R+D_P+D_D$.
\end{itemize}

\item The square-wave (sq.w.) approximation of an average superoutburst, as described by \cite{2011Patterson}. 
Although its parameters are often very similar to actual amplitude and duration of superoutbursts, we decided to keep them in the catalog, since they are much easier to measure. Thus, the square wave approximation provides even 2-3 times more data points as compared to $D_{SO}$, which requires measurements of three other parameters.
\begin{itemize}
\item $V_{sq.w.}$ -- apparent V magnitudes of the square-wave,
\item $D_{sq.w.}$ -- duration of the square-wave.
\end{itemize}

\item Minimal and maximal values of normal cycle length, $P_c$, that is a time interval between two subsequent short outbursts for U~Gem and Z~Cam stars, or two subsequent normal outbursts for SU~UMa stars.
\item Minimal and maximal values of supercycle length, $P_{sc}$, that is a time interval between two subsequent long outbursts for U~Gem and Z~Cam stars, or two subsequent superoutbursts for SU UMa stars.
\item Presence of a precursor outburst before superoutburst.
\end{enumerate}

These parameters differ a little bit for Z~Cam stars. To maintain the consistency with the catalog of \cite{2003Ritter} we recorded the brightness of standstills for these stars as their "Mag4" which corresponds to our "$V_{SO}$". Their maximum brightness of long outbursts is recorded as $V_{sq.w.}$, its duration as $D_{sq.w.}$, and maximum brightness of short outbursts as $V_{NO}$. 

The photometric features enumerated so far were directly noted or measured from light curves.  Additionally, we chose to study the following parameters measured indirectly.

\begin{enumerate}
\item  Amplitudes of all types of outbursts, defined as follows:

\begin{itemize}
\item $A_{NO}$ -- the amplitude of normal outbursts of SU~UMa stars or short outbursts of U~Gem and Z~Cam stars, that is the difference of the  apparent V magnitude levels at $V_{NO}$ and $V_{min}$, 
\item $A_{SO}$ -- the amplitude of superoutbursts of SU~UMa stars or long outbursts of U~Gem and Z~Cam stars, that is the difference of the  apparent V magnitude levels at $V_{SO}$ and $V_{min}$ for SU~UMa and U~Gem stars and $V_{sq.w.}$ and $V_{min}$ for Z~Cam stars,
\item $A_{sq.w.}$ -- the amplitude of the square wave approximation of superoutbursts or long outbursts, defined as the difference of the  apparent V magnitude levels at $V_{sq.w.}$ and $V_{min}$ for SU~UMa and U~Gem stars and $V_{NO}$ and $V_{min}$ for Z~Cam stars. 
\end{itemize}

\item Rates of brightness changes during three phases of superoutbursts of SU~UMa stars or long outbursts of U~Gem and Z~Cam stars:
\begin{itemize}
\item $\tau_{r}$ -- rise rate, i.e., $\frac{D_R}{A_{SO}}$, 
\item $\tau_{d,pl}$ -- decline rate of plateau, i.e.,  $\frac{D_P}{A_{SO}-A_{sq.w.}}$, 
\item $\tau_{d}$ -- decline rate, i.e.,  $\frac{D_D}{A_{sq.w.}}$. 
\end{itemize}

\item Mean value of $P_c$ \& $P_{sc}$ calculated from minimum and maximum values recorded in the catalog. 
\end{enumerate}

For objects in which light curves exist of multiple outbursts, the determination of parameters was not based on an aggregate of those outbursts, as it is mostly done in a single object analysis. Such an approach for hundreds of light curves is far beyond our capabilities. Instead, we manually measured these parameters based on as complete light curves as possible, and recorded an average value of each of the parameters. In case of a strong variability of a basic outbursts parameter, we favoured the observations taken during the recent years or decades over older ones because of the improving quality of the photometric data coming from amateur observers. 

Our measurements do not give a specific and accurate values of errors for each object, however, we can estimate a mean precision of every parameter (see Section~\ref{sec-catalog}). 
We have reached an adequate precision for our purposes, since based on for example the cycle length vs. supercycle length relation presented in Section~\ref{PscPc}, we were accurate enough to reconstruct very precisely some of the known relations. Every other parameter in the catalogue was recorded in the same, very careful and exact way as these cycle and supercycle lengths. In case of any doubts while determining a value of a parameter, we did not record anything at all.

\subsection{Light curves}

\begin{figure}
\centering
\includegraphics[width=0.5\textwidth]{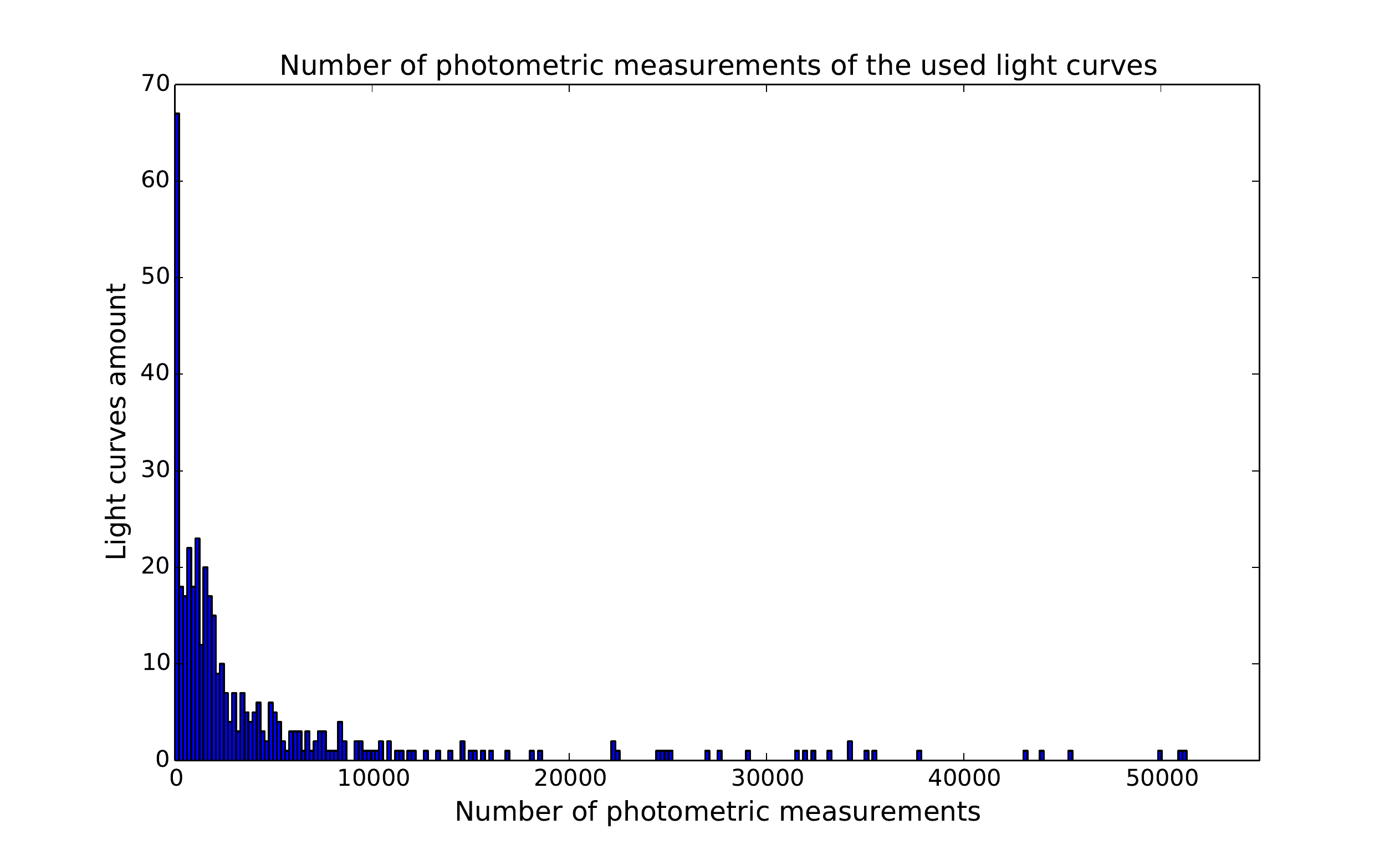}
\caption{Number of photometric measurements of the used light curves. The \textit{X} axis is truncated at 55000 for a better visibility, although there are additional 11 objects with > 60000 observations with maximum of 1447957 observations for the Kepler light curve of V1504~Cyg.} 
\label{fig-LCamount}
\end{figure}

\begin{figure}
\centering
\includegraphics[width=0.5\textwidth]{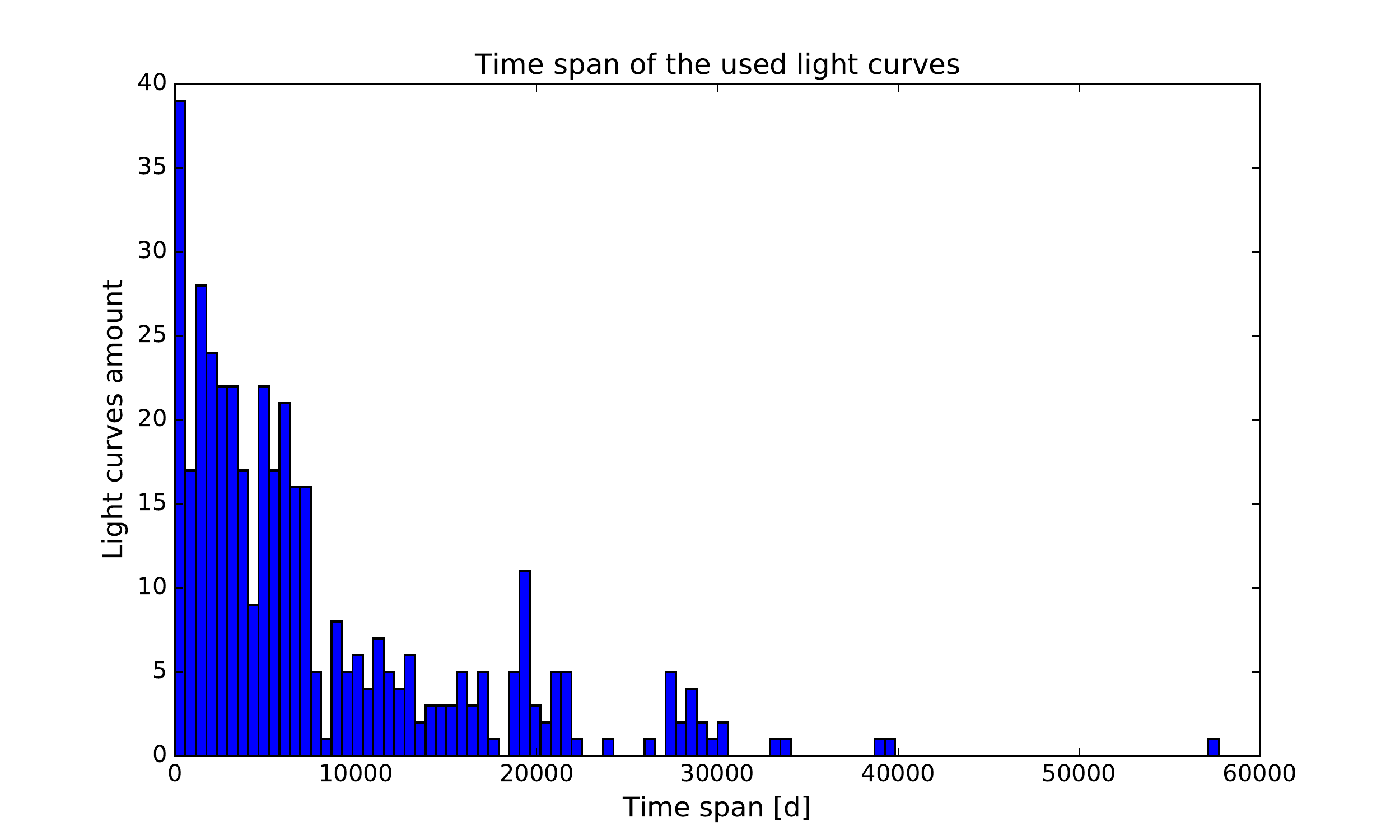}
\caption{Distribution of the time span of the light curves used in the catalogue. The maximum of over 150 years of observations corresponds to U~Gem.} 
\label{fig-timespan}
\end{figure}

While describing such a rich set of photometric data, it is not possible to characterise each of the light curves separately. However, in Fig.~\ref{fig-LCamount} and \ref{fig-timespan} we show a cumulative characteristic of the whole data set in terms of the amount of photometric measurements and time span of the light curves used to determine the parameters given in our catalogue.

Typical cadences of the data over the long term vary from one object to another and sometimes even vary for different epochs of observations of the same star. 
However, the cadences variations do not influence the parameters determinations at all. Even in case of gaps in the light curves, we definitely did not miss any outburst while determining the cycles length. In case of any doubts or a possibility of a not sufficient time coverage of observations, we did not record any parameter. If there is a record in the catalogue, we are convinced that we were able to measure it certainly. 

The same is true for the sampling cadence over the outbursts themselves and determination of their rises and declines. For such a big data set coming from such a diverse data sources, the sampling cadences vary significantly. To measure the specific parameters each time we used only the parts of the light curves where rises and declines of outbursts were clearly resolved in the photometric data. We recorded a parameter only when it was certain. 
Based on data covering years of observations, it was possible to catch a star some times during a rise of a superoutburst and other times during superoutburst decline. 
That is why, even for cases of a poor time coverage of light curves, different parameters have been measured based on different parts of observations. 

\begin{figure*}
\centering
\includegraphics[width=0.9\textwidth]{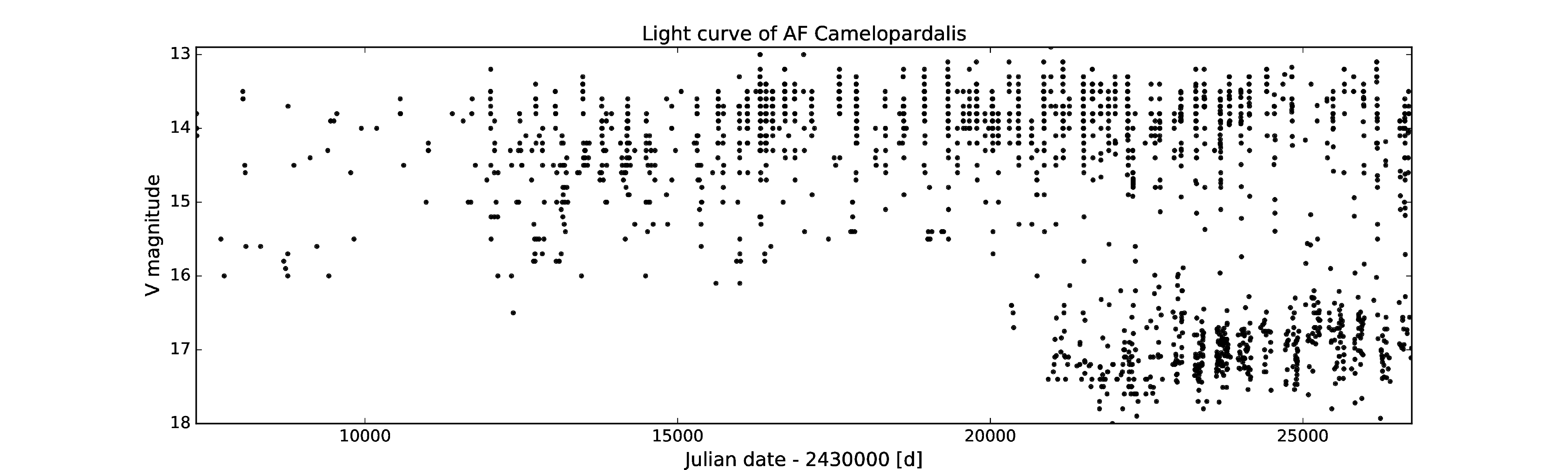}
\includegraphics[width=0.9\textwidth]{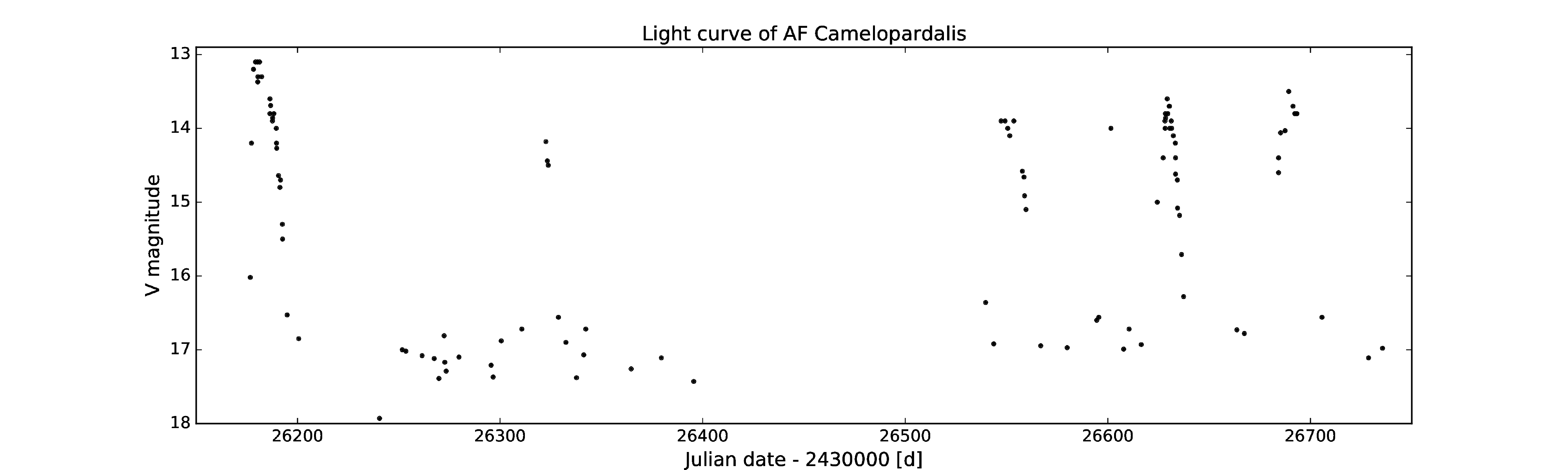}
\caption{A poor quality, but still useful, light curve of AF~Cam.
Top: the whole available light curve of AF~Cam where the oldest observations reached only the brightest magnitude levels during outbursts.
Bottom: a zoom-in of the light curve where a few outbursts can be noticed.} 
\label{fig-AFCam1}
\end{figure*}

\begin{figure*}
\centering
\includegraphics[width=0.9\textwidth]{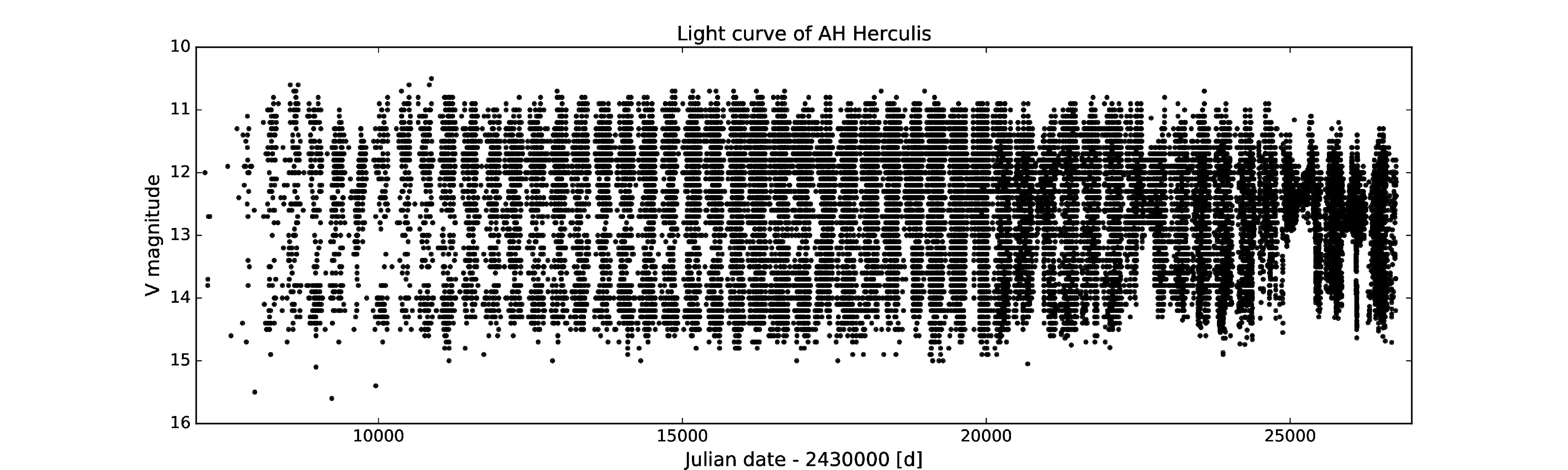}
\includegraphics[width=0.9\textwidth]{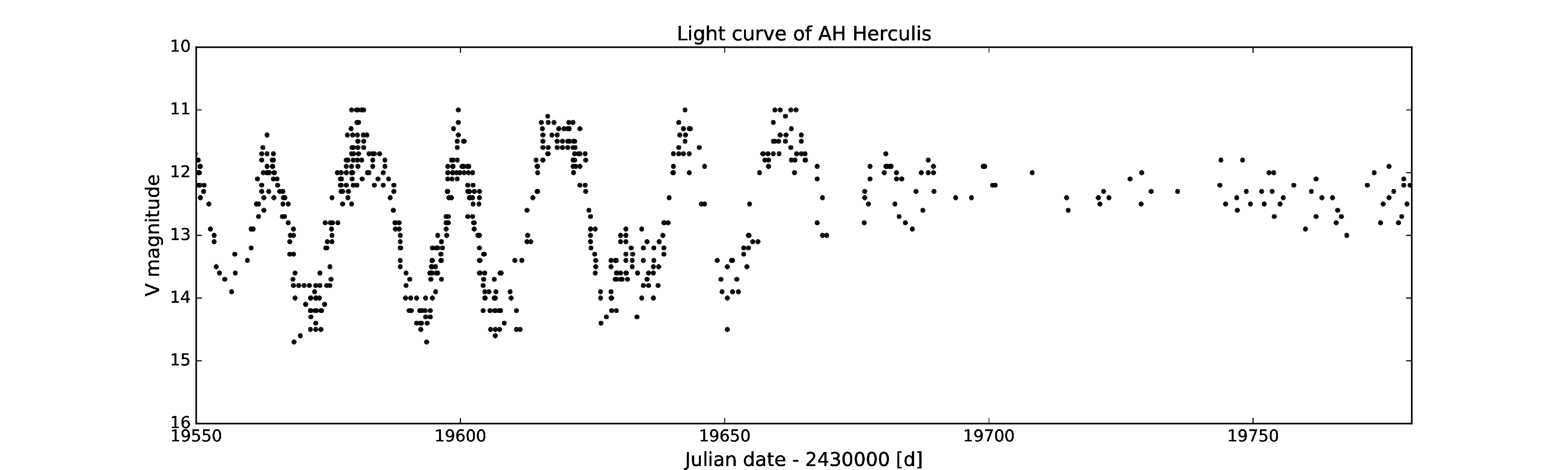}
\caption{An average quality light curve example. 
Top: the whole available light curve of a Z~Cam-type dwarf nova, AH Her. 
Bottom: a zoom-in of the light curve where a standstill is nicely visible after a number of outbursts.} 
\label{fig-AHHer}
\end{figure*}

\begin{figure*}
\centering
\includegraphics[width=0.9\textwidth]{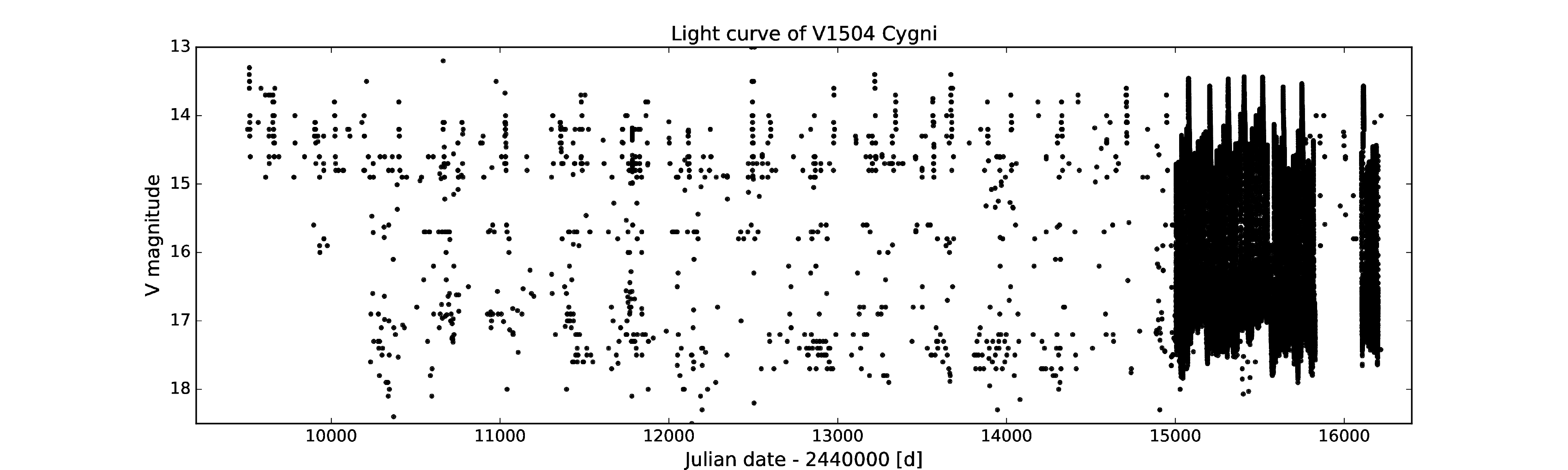}
\includegraphics[width=0.9\textwidth]{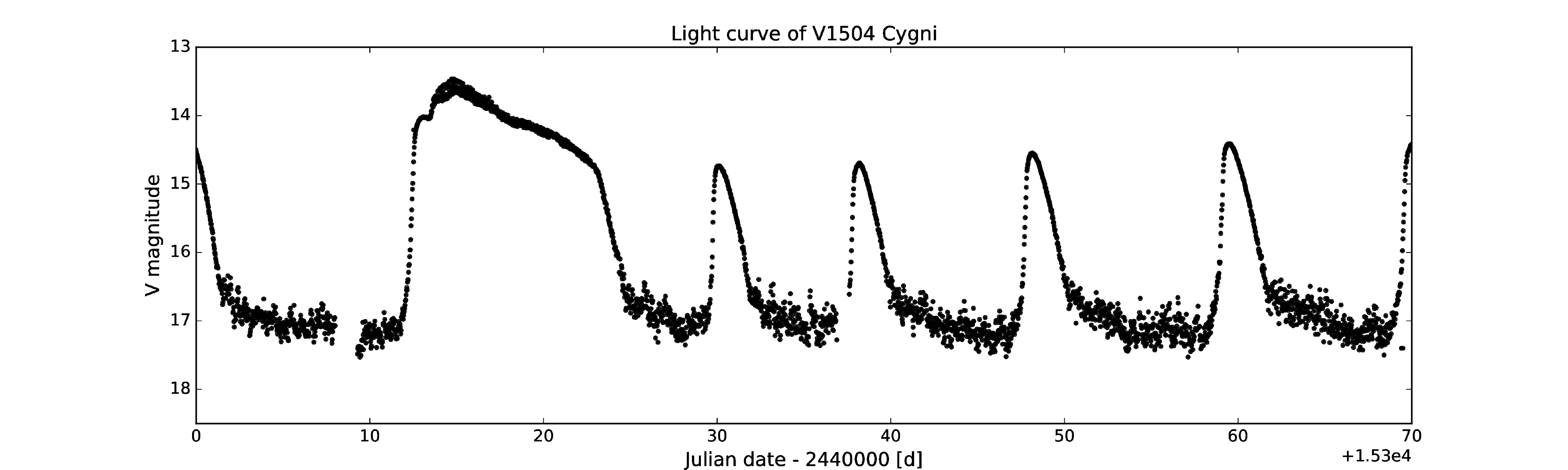}
\caption{The best quality light curve from the Kepler satellite. 
Top: the AAVSO light curve together with the Kepler light curve of V1504~Cyg. 
Bottom: a zoom-in of the Kepler light curve.} 
\label{fig-kepler}
\end{figure*}

To give an impression of the actual quality of the data which was used to measure the basic brightness levels and outbursts duration and frequency, we show some examples in Figures~\ref{fig-AFCam1}, \ref{fig-AHHer}, and \ref{fig-kepler}. We chose these examples so that a comparison with the first rows of the catalogue values (Table~\ref{tab-firstRows}) is possible for poor and average quality data. The best quality data did not appear within the first catalogue's rows.

Examples of poor quality light curves from objects in the Table~\ref{tab-firstRows} are AF~Cam and AK~Cnc. We show the first of them in Figure~\ref{fig-AFCam1}. The total light curve in this case covers about 200 outbursts but only about 30\% of them can be actually used in the analysis due to poor sampling cadences. 

An example of the average quality of photometric data is shown in Figure~\ref{fig-AHHer}, presenting the light curve of a Z~Cam-type dwarf nova, AH~Her. 
The total light curve of this object covers about 500 outbursts. Their amplitudes differ from each other, especially while comparing distant observations. In such a case the most recent (and thus more precise) data was definitive. 

The best quality data comes of course from the Kepler satellite. We show such example in Figure~\ref{fig-kepler}, where the AAVSO and Kepler light curves are shown together for V1504~Cyg. Such exquisite cadence sampling of particular outbursts is available only for a few stars in our sample. In this case there are less than a hundred outbursts observed by Kepler and a few hundreds of them, with worse cadence sampling, from AAVSO.

When it comes to the photometric precision of the data, is it worth mentioning that most of the light curves used in the analysis come from amateur observers. They mostly provide data of a good quality, however, the final light curves are composed of many different observations sometimes with a biased calibration. That is why we need to account for an uncertainty in this matter. 
For the apparent brightness $V$ of any type we estimate this error at the level of 0.1~mag, which is the upper limit of this uncertainty, because amateur observers have often a much better precision.
The final error of our \textit{V} measurements, taking into account heterogeneity of the outbursts shapes within a particular light curve of a single object, we estimate at 0.2~mag.

All of the light curve data is publicly available on the websites given in Section~\ref{data-sources}. The only exception is the data from our papers listed there as the last item. Thus, we make this data also available on the same web location as the catalogue, http://users.camk.edu.pl/magdaot/DN.html.

\subsection{The final catalog}
\label{sec-catalog}

The final catalog contains photometric features of DN from catalogs of \cite{2001Downes}, \cite{2003Ritter}, and \cite{2011Patterson}, listed in Sect.~\ref{data-sources} together with our measurements described in Sect.~\ref{phot-feat}.
It contains 653 dwarf novae, with the distribution shown in Table~\ref{tab-stat}. 

\begin{table} 
\caption{Number of DN in our catalog.}
\begin{tabular}{@{}ccc@{}}
\hline 
class & number of objects & percentage  \\ 
\hline 
all DN & 653 & 100\% \\ 
SU UMa & 459 & 70\% \\ 
U Gem & 88 & 13\% \\ 
Z Cam & 30 & 5\% \\ 
\hline 
SU UMa subclass & number of objects & percentage  \\ 
\hline 
ER UMa & 11 & 2\% \\ 
SU UMa & 386 & 59\% \\ 
WZ Sge & 62 & 9\% \\ 
\hline 
unknown type & 76 & 12\% \\
\hline
\end{tabular} 
\label{tab-stat}
\end{table}

Table~\ref{tab-firstRows} presents the first several rows of the final catalogue containing all of the available information. Each row consists of 32 columns coming from the used catalogues or containing values directly measured from the light curves. 

The catalogue contains the following data and estimates:
\begin{itemize}
\item \textit{Name} = the GCVS name of the dwarf nova,
\item \textit{RA} = Right Ascension,
\item \textit{Dec} = declination,
\item \textit{type(1)} = Type2 from the catalogue of \cite{2003Ritter} (its Type1 was "dwarf nova"),
\item \textit{type(2)} = Type3 from the catalogue of \cite{2003Ritter},
\item \textit{type(3)} = VarType from the catalogue of \cite{2001Downes}, for comparison, 
\item \textit{V$_{min}$, V$_{NO}$, V$_{SO}$, V$_{sq.w.}$, D$_{NO}$, D$_{R}$, D$_{P}$, D$_{D}$, D$_{sq.w.}$, P$_{sc}$, P$_{sc}^{max}$, P$_{c}$, P$_{c}^{max}$}, as described in Section\ref{phot-feat},
\item \textit{pre} = precursor's presence,
\item \textit{P$_{orb}$ } = orbital period in hours
\item \textit{uP$_{orb}$}  = uncertainty flag for orbital period,
\item \textit{P$_{sh}$ } = superhump period,
\item \textit{uP$_{sh}$ } = uncertainty flag for superhump period,
\item \textit{incl } = inclination,
\item \textit{eincl } = error of inclination,
\item \textit{M1/M2 } = mass ratio from the catalogue of \cite{2003Ritter}, 
\item \textit{eM1/M2 } = error of M1/M2,
\item \textit{q=M2/M1 } = inverse of the mass ratio M1/M2, as it is more common in CVs' analysis,
\item \textit{qJP } = mass ratio from \cite{2011Patterson},
\item \textit{qCalc } = mass ratio calculated from the superhump period excess with the use of Eq.~\ref{eq:epsilon22} and \ref{eq:epsilon222}.
\end{itemize}

Uncertainties and errors of the parameters are given when available, which is unfortunately not often. However, we can estimate the uncertainties of the rest of the parameters, based on the light curve analysis. 

For the apparent brightness of any type (all \textit{Vs}) we estimate the error at the level of 0.2~mag, as mentioned in the previous subsection. This is the typical accuracy of the light curves in magnitudes, taking into account the calibration differences coming sometimes from diverse observers and heterogeneity of the outbursts shapes within a light curve of a single object. 

When it comes to durations of specific parts of outbursts of any type (all $D$s), we estimate their precision at the level of 0.5~day, which is the upper limit of the error for all the light curves and also accounts for the heterogeneity of the outbursts shapes within a light curve of a single object. In the case when we had a photometric data but could not measure some \textit{D} with such an accuracy, we did not record this parameter into the catalogue at all. 

Cycle and supercycle lengths (\textit{Ps}) are measured with the same accuracy as \textit{Ds}, however, the heterogeneity in this matter was more prominent. Most of the time the cycle's lengths vary and the recorded \textit{Ps} are mean values of cycle and supercycle lengths.

Periods taken from the known catalogues most often come without precise uncertainties. A literature review would probably allow to collect some of them, however, since it is very time-consuming, we did not do it for this first version of our catalogue. We plan to improve this issue in our future catalogues releases. 

Mass ratio in the column "M1/M2" comes from \cite{2003Ritter} with errors given in the next column. 
\textit{qJP} are also given without uncertainties. Most of them come from the superhump period excess, with several examples from eclipses and radial-velocity measurements. 
\textit{qCalc} is our calculation of the mass ratio based on the superhump period excess. This relation is least accurate for small period excess values.

\begin{landscape}
\begin{table}
\centering 
\caption{Our final catalogue containing photometric features of DN from catalogues 
listed in Sect.~\ref{data-sources}, together with our measurements described in Sect.~\ref{phot-feat}.
We show only the first several rows of the catalogue. They are divided into two parts, since the original rows were too long to fit into the page. 
The complete table containing 653 dwarf novae is available at http://users.camk.edu.pl/magdaot/DN.html, together with an explanation of each column, which is also given in Section~\ref{sec-catalog} of this paper.}
\begin{tabular}{|l|l|l|l|l|l|r|r|r|r|r|r|r|r|r|r|r|r|r|r|r|l|r|l|l|r|l|r|r|l|r|r|r|r|r|r|l|}
\hline
  \multicolumn{1}{|c|}{Name} &
  \multicolumn{1}{c|}{RA} &
  \multicolumn{1}{c|}{Dec} &
  \multicolumn{1}{c|}{type(1)} & 
  \multicolumn{1}{c|}{type(2)} & 
  \multicolumn{1}{l|}{type(3)} & 
  \multicolumn{1}{c|}{V$_{min}$} &
  \multicolumn{1}{c|}{V$_{NO}$} &
  \multicolumn{1}{c|}{V$_{SO}$} &
  \multicolumn{1}{c|}{V$_{sq.w.}$} &
  \multicolumn{1}{c|}{D$_{NO}$} &
  \multicolumn{1}{c|}{D$_{R}$} &
  \multicolumn{1}{c|}{D$_{P}$} &
  \multicolumn{1}{c|}{D$_{D}$} &
  \multicolumn{1}{c|}{D$_{sq.w.}$} &
  \multicolumn{1}{c|}{P$_{sc}$} &
 \\
\hline
AB Dra & 19 49 06.5 & +77 44 24 & ZC &  & UGZ   & 14.5 & 12.5 & 13.8 & 12.2 & 5.0 &     &     &     & 10.0 & 50.0   \\
AB Nor & 15 49 15.5 & -43 04 49 & SU &  & UGSU  & 20.3 & 14.7 & 13.9 & 14.6 &     &     &     &     & 15.0 &        \\
AC LMi & 10 19 47.3 & +33 57 54 &    &  &       & 17.5 & 15.1 &      &      &     &     &     &     &      &        \\
AD Men & 06 04 30.9 & -71 25 23 & SU &  & UGSS  & 17.6 &      & 14.0 &      &     &     &     &     &      & 320.0  \\
AF Cam & 03 32 15.6 & +58 47 22 & UG &  & UG    & 17.0 & 13.4 & 13.0 & 13.2 & 9.0 & 2.0 & 9.0 & 7.0 & 11.0 & 80.0   \\
AG Hya & 09 50 29.8 & -23 45 17 & UG &  & UG    & 19.2 & 14.5 &      &      & 5.0 &     &     &     &      &        \\
AH Eri & 04 22 38.1 & -13 21 30 & UG & IP & UG/DQ & 17.7 & 13.8 &    &      & 7.0 &     &     &     &      &        \\
AH Her & 16 44 10.0 & +25 15 02 & ZC &  & UGZ   & 14.3 & 11.2 & 12.4 & 11.3 & 9.0 & 1.0 & 8.0 & 4.0 & 9.0  &        \\
AK Cnc & 08 55 21.2 & +11 18 15 & SU &  & UGSU  & 18.7 &      & 12.8 & 13.9 &     & 1.0 & 11.0& 9.0 & 12.0 & 200.0  \\
... & ... & ... & ... & ... & ...  & ... &  ... & ... & ... & ... & ... & ... & ... & ... & ...  \\
\hline
\end{tabular} 
\label{tab-firstRows}
\end{table}

\begin{table}
\centering 
\contcaption{Our final catalogue containing photometric features of DN from catalogues 
listed in Sect.~\ref{data-sources}, together with our measurements described in Sect.~\ref{phot-feat}.
We show only the first several rows of the catalogue. They are divided into two parts, since the original rows were too long to fit into the page. 
The complete table containing 653 dwarf novae is available at http://users.camk.edu.pl/magdaot/DN.html, together with an explanation of each column, which is also given in Section~\ref{sec-catalog} of this paper.}
\begin{tabular}{|l|l|l|l|l|l|l|r|r|r|r|r|r|r|r|r|r|r|r|r|r|l|r|l|l|r|l|r|r|l|r|r|r|r|r|r|l|}
\hline
  \multicolumn{1}{|c|}{Name} & 
  \multicolumn{1}{c|}{...} &
  \multicolumn{1}{c|}{P$_{sc}^{max}$} &  
   \multicolumn{1}{c|}{P$_{c}$} &
  \multicolumn{1}{c|}{P$_{c}^{max}$} &
  \multicolumn{1}{c|}{pre} &
  \multicolumn{1}{c|}{P$_{orb}$} &
  \multicolumn{1}{c|}{uP$_{orb}$} &
  \multicolumn{1}{c|}{P$_{sh}$} &
  \multicolumn{1}{c|}{uP$_{sh}$} &
  \multicolumn{1}{c|}{incl} &
  \multicolumn{1}{c|}{eincl} &
  \multicolumn{1}{c|}{M1/M2} &
  \multicolumn{1}{c|}{eM1/M2} &
  \multicolumn{1}{c|}{q=M2/M1} &
  \multicolumn{1}{c|}{qJP} &
  \multicolumn{1}{c|}{qCalc} \\
\hline
  AB Dra & ... &  	 & 11.0 &     	&  & 0.152 		&   &  			&   &   		&  	&   	&  		&  	&		&   \\
  AB Nor & ... &  	 &  	  &     	&  & 0.077 		& * & 0.0796 	&   &   60.0   	&  	&  		&   	&  	&    	& 0.156    \\
  AC LMi & ... &  	 & 	  &     	&  & 0.0794 	&   &  			&   &   		&  	&   	&   	&  	&  		&    \\
  AD Men & ... &  	 &  	  &     	&  & 0.0922 	&   & 0.0966 	&   &   		&   &  		&   	&   & 0.22 	& 0.221  \\
  AF Cam & ... & 230.0 & 60.0 & 80.0    &  & 0.324078 	&   &  			&   &   		&  	&    	&   	&  	&     	&  \\
  AG Hya & ... &  	 & 15.0 & 20.0   	&  & 0.238 		&   &  			&   &   		&  	&   	&   	&  	&  		&      \\
  AH Eri & ... &  	 & 60.0 & 80.0   	&  & 0.2391 	&   &  			&   &   		&  	&   	&  		&  	&  		&     \\
  AH Her & ... &  	 & 12.0 & 16.0 	&  & 0.258116 	&   &  			&   &   46.0   & 3.0 & 1.25 & 0.08 	& 0.8 &   	&   \\
  AK Cnc & ... & 350.0&      &     	&  & 0.0651 	&   & 0.0674 	&   &   40.0   &  	&   	&  		&  	& 0.16 	& 0.164   \\
  ...	 & ... &  ...	 & ...  & ... 	&...  & ... 	& ...  & ...&...   &...   &...  &...   	&...	& ... 	& ... 	& ...   \\

\hline
\end{tabular}
\end{table}
\end{landscape}

\section{Relations}
\label{sec-relations}

This section gives a number of correlations of parameters of interest. Some of them are tight and explicit, others are less clear, but may provide suggestions on apparent trends in the data. 

The description of the statistical method used to decide on the existence of a correlation between two given parameters is presented in Sect.~\ref{method}.
In Sections~\ref{najwazniejsza}, \ref{najwazniejsza2}, and \ref{najwazniejsza3} we discuss in our opinion the most important relations.  
Section~\ref{exam-corr} gives an overview of a few other correlations. In the end, the Appendix contains remaining dependencies. 

In each case concerning superoutbursts, only data points of SU~UMa-type stars were used to find a fit. However, in the scatter plots we also show data for U~Gem and Z~Cam stars, for comparison.
This may raise concerns on the correctness of such approach in the context of the TTI model which suggests different physical cause behind normal and superoutbursts. 
We are aware of these physical  differences, but we decided to show all dwarf novae in such plots for comparison. 
For instance, the pure thermal instability model does not distinguish superoutburst from other long outbursts of long-period systems. We do not intent to favour any of the existing theories.

We kept the same symbols and colors for representatives of each DN subclass throughout the whole work, so they are easy to compare. The number of data points vary from one relation to another because not all parameters were measured for all objects. Our catalog is as complete as possible, but unfortunately there is still a lot of data missing. Thus, we treat this work as the first approach to a complete statistical study and we hope to collect more and/or better light curves in the future to improve the statistics.

\subsection{Method}
\label{method}

We examined all the possible parameters listed in the previous Section in order to verify if they are correlated. The most interesting of them are presented here, whereas the rest is given in the Appendix in a more complex form.

Based on \cite{2012Wall}, for each relation we calculated the \textit{Pearson product-moment correlation coefficient} (sample correlation coefficient):
\begin{equation}
r = \frac{\sum^N_{i=1}(X_i-\overline{X})(Y_i-\overline{Y})}{\sqrt{\sum^N_{i=1}(X_i-\overline{X})^2 \sum^N_{i=1}(Y_i-\overline{Y})^2}}
\end{equation}
which is a measure of the linear correlation between two variables in statistics. 
It gives a value between $+1$ and $-1$ inclusive, where $1$ is total positive correlation, $0$ is no correlation, and $-1$ is total negative correlation. 
To evaluate the presence or absence of each correlation we used the classical methodology of hypothesis testing. 
We selected:
\begin{itemize}
\item the null hypothesis ($H_0$) = no correlation,
\item the alternative hypothesis ($H_1$) = a presence of a correlation,
\item level of significance, i.e., the maximum probability of rejecting the null hypothesis when it is true,  $\alpha=5\%$.
\end{itemize}
To test the significance of a non-zero value for $r$, we used statistical tables with critical values of the Pearson product-moment correlation coefficient.

Additionally, we calculated the \textit{coefficient of determination}, $R^2$. Its value may be interpreted as percent of the variation in the response variable that can be explained by the explanatory variables. The remaining percent can be attributed to unknown variables or inherent variability \citep{2012Wall}.

\subsection{Stolz and Schoembs Relation}
\label{najwazniejsza}

Probably the most interesting correlation derived from our sample is the Stolz and Schoembs relation that we already presented during the 19th European Workshop on White Dwarfs \citep{2015ASPC..493..517O}, however, we mention it here for completeness. 

For the first time \cite{1984Stolz} showed an observational evidence that the orbital period-superhump period excess, $\epsilon$, and the orbital period, $P_{orb}$, are correlated, while \cite{2001Patterson} proposed: 
\begin{equation}
\epsilon = \frac{\Delta P}{P_{orb}} = \frac{P_{sh}-P_{orb}}{P_{orb}}
\label{eq:epsilon22}
\end{equation}
and
\begin{equation}
\epsilon = 0.216(\pm~0.018)~q,
\label{eq:epsilon222}
\end{equation}
where $q$ is the mass ratio.

This is a very useful relation, since it provides a plane to trace evolution of dwarf novae. It can also help to estimate typical values of the mass ratio for a given orbital period and thus, with known donors' masses, to set boundaries for theoretical models \citep{2005Patterson}.

\begin{figure}
\centering
\includegraphics[width=0.45\textwidth,angle=270]{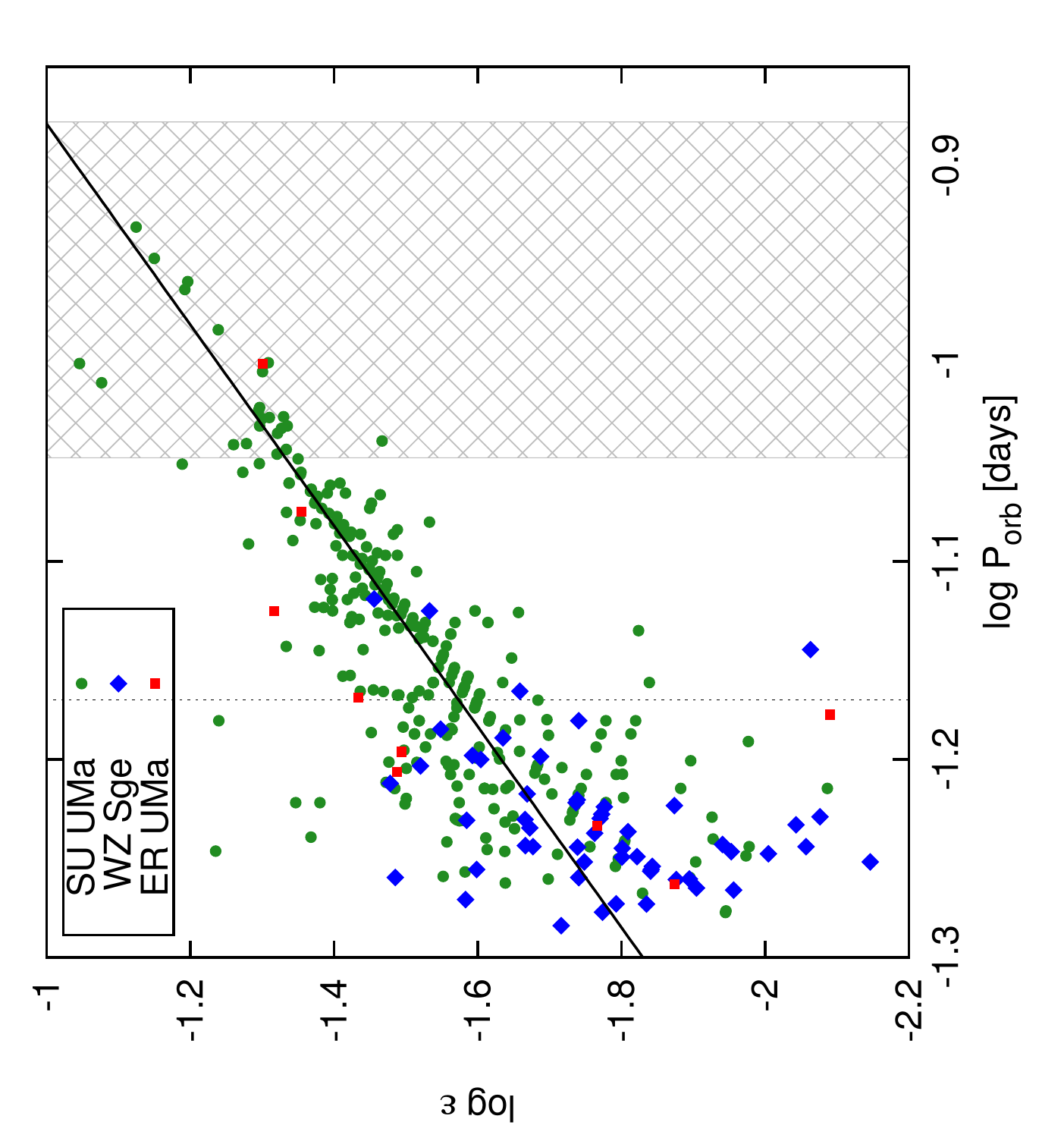}
\caption{The Stolz and Schoembs relation, i.e., relation between the orbital period-superhump period excess and the orbital period for all DN in our catalog. The solid line shows a fit for objects with log $P_{orb}>-1.17$, that is where the relation is distinct. The hatch pattern presents the period gap. The vertical line marks the place of departure from the fit and dispersion increase.}
\label{fig-SSrel}
\end{figure}

The Stolz and Schoembs relation for all dwarf novae in our sample for which both periods ($P_{orb}$ and $P_{sh}$) are known, and thus $\epsilon$ could be calculated, is presented in Figure~\ref{fig-SSrel}. 
It consists of many superhumpers which are arranged in a surprisingly beautiful and distinct line corresponding to the path of a DN during its evolution from long to short orbital periods. 
Below $\log~P_{orb}$~[d]~$=-1.17$ this path is not so clear anymore, since there are two separate paths in this period range, i.e., the continuation of evolution towards the minimum period as well as in the opposing direction by period bouncers.
Thus, data points with log $P_{orb}>-1.17$, that is where the relation is distinct, were used to find a corresponding fit of a linear function which in the Figure is shown with the solid line. We found:

\begin{equation}
\log~\epsilon = 1.97(0.10) \cdot \log~P_{orb} \rm{[d]} + 0.73(0.11)
\label{fit-epsilon}
\end{equation}

In spite of the fact that the objects present in this plot are only superhumpers, that is objects below the period gap, the fit seems to work also for these CVs which are located inside the gap. For a typical DN, created above the gap, the mass ratio and thus also $\epsilon$ do not change during the evolution through the period gap \citep{2011Knigge}.
The relation presented by Eq.~\ref{fit-epsilon} works there only for these rare SU~UMa stars which were created in the period gap. 
It is worth mentioning that this relation is not valid for novalike objects and long-period systems above the gap which show superhumps, e.g. U~Gem or TV~Col \citep{2004SmakSmak, 2003Retter} and at higher mass ratios this relation becomes non-linear. However, the existence of superhumpers in this orbital period range is a challenge to the theory. 

As we showed in \cite{2015ASPC..493..517O}, donor stars' masses above and below the period gap, based on our data, are the following. 
Taking into account the period gap boundaries of \cite{2006Knigge}, namely 
$P_{gap, -} = 2.15 \pm 0.03$~h, and $P_{gap, +} = 3.18 \pm 0.04$~h, 
and a typical white dwarf mass, $M_1 = 0.6 M_\odot$, we found the following masses of a donor star:
\begin{itemize}
\item above the period gap: $M_{2,+} = 0.28 \pm 0.09 M_\odot$, 
\item below the period gap: $M_{2,-} = 0.13 \pm 0.05 M_\odot$, 
\item and at $\log P_{orb}=-1.17$, where the relation becomes distinct: $M_{2,d} = 0.07 \pm 0.03 M_\odot$.
\end{itemize}

Donor stars of dwarf novae immediately flanking the period gap, so the value of $M_{2,+}$ gives the upper limit for the rare objects created in the period gap. 
These estimated values are in agreement with recent theoretical models. 
For instance, \cite{2011Knigge} showed that donors' mass for CVs in the period gap, according to the revised standard model of CVs' evolution, is $M_{2} = 0.20\pm0.02 M_\odot$. 

For comparison, we calculated also the same values only this time for a heavier white dwarf of $M_1 = 0.8 M_\odot$. The results are the following:
\begin{itemize}
\item $M_{2,+} = 0.37 \pm 0.12 M_\odot$, 
\item $M_{2,-} = 0.17 \pm 0.06 M_\odot$, 
\item $M_{2,d} = 0.10 \pm 0.04 M_\odot$. 
\end{itemize}
Thus, some diversity of masses of white dwarfs could explain the observed scatter in the diagram.

\subsection{Kukarkin-Parenago Relation}
\label{najwazniejsza2}

The Kukarkin-Parenago Relation (KPR) is one of relations with the longest tradition in the field of dwarf novae studies \citep{2003Warner}:

\begin{equation}
A_{NO} [mag] = 0.70(0.43) + 1.90(0.22)\cdot{\log} P_c [d]
\label{eq-KPR}
\end{equation}
It connects the amplitude of dwarf novae outbursts with their recurrence time, given in magnitudes and days, respectively. 
It was originally presented by \cite{1934Kukarkin} and later on by a few other scientists {(e.g. \citep{1976Kholopov}.

Although KPR is used by many authors, it has been also questioned in the past, for instance by \cite{1977PayneGaposchkin}. 

In turn \cite{1985vanParadijs} found the following relation
\begin{equation}
A_{NO} [mag] = 1.84(0.44) + 1.40(0.23)\cdot{\log}P_c [d]
\label{eq-KPR2}
\end{equation}
However, in Fig.~1 of his publication the scatter of points is extensive and it is difficult to see a correlation there. The only reason to claim otherwise is the value of the correlation coefficient implying that the relation is real. In any case, it is ambiguous.

\cite{1987Warner} showed that to tighten the relation $A_{NO}$ needs to be freed from distortions from the secondary and the hot spot, i.e., it must be calculated from the minimum light coming from only the disk.
The influence of the hot spot can be estimated by the use of orbital variations of a light curve and the contribution from the secondary in the visual band can be  neglected.
Unfortunately, we are not able to directly compare our data with his result because, based mostly on amateur observations, we cannot accurately estimate the minimum light from the disk alone. It is beyond our precision. Nevertheless, the corrections of amplitudes used by \cite{1987Warner}, i.e. $\Delta$ shown in his Table~1, are most of the time equal to zero, for several stars equal to 0.2~mag, and the maximum value is 0.5~mag for one single object. Thus, the $\Delta$ corrections are not significant, and we claim that a comparison to our data is reasonable in this case. 

Figure~\ref{fig-Pc-ANO} shows no significant Kukarkin-Parenago Relation for our data. 
\begin{figure}
\centering
\includegraphics[width=0.5\textwidth]{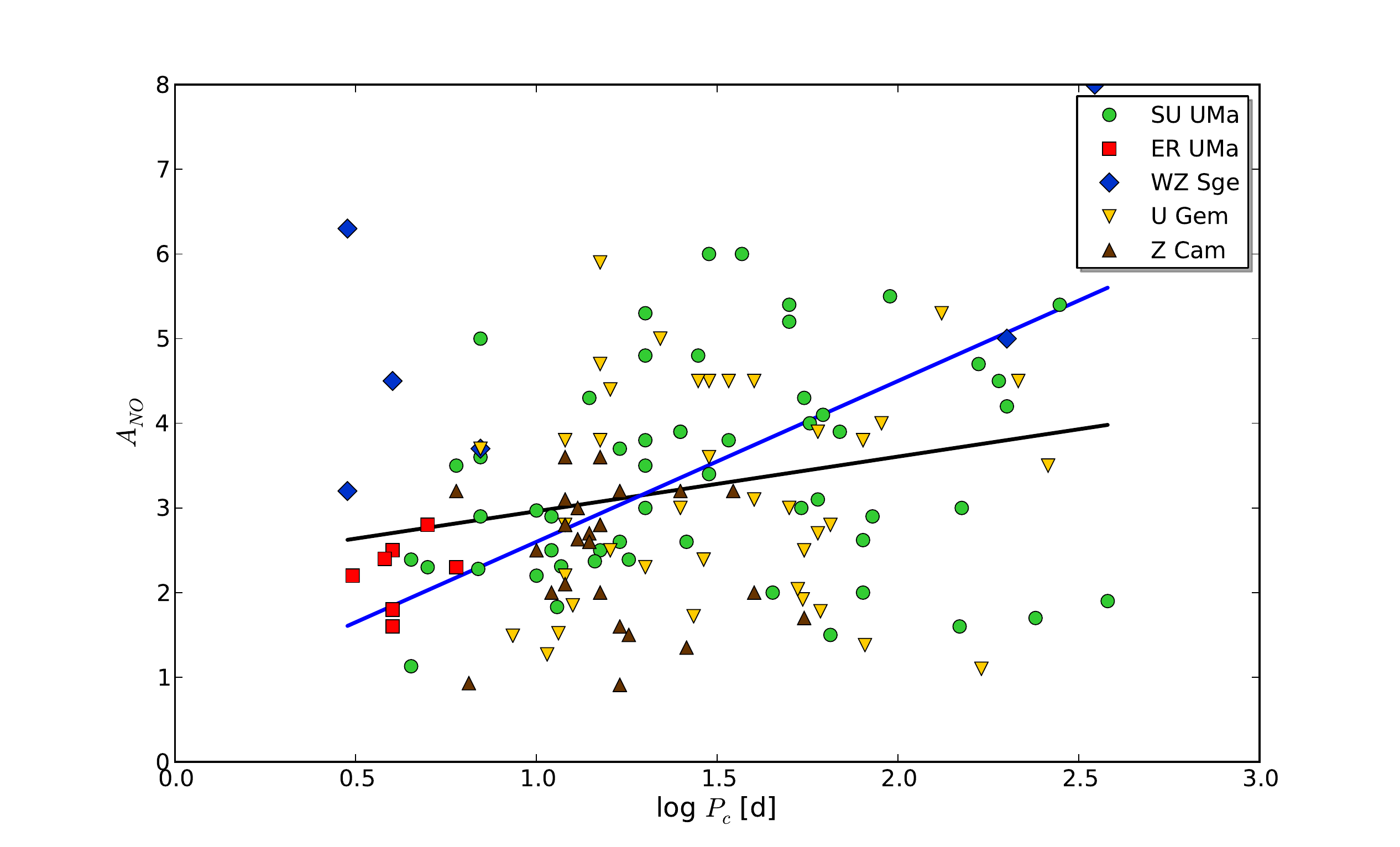}
\caption{The lack of the Kukarkin-Parenago Relation for our data. The black line presents the formal linear fit, whereas the blue line stands for the commonly accepted and used KPR (Eq.~\ref{eq-KPR}).}
\label{fig-Pc-ANO}
\end{figure}
The black line presents the formal linear fit:
\begin{equation}
A_{NO} [mag] = 2.31(0.37)+0.64(0.24)\cdot{\log}P_c [d],
\label{eq-KPR3}
\end{equation}
whereas the blue line stands for the commonly accepted and used KPR (Eq.~\ref{eq-KPR}). 

Even though the correlation coefficient for our fit gives a value ($r=0.23$ for N=127 data points) indicating that the relation is real, we claim that the existence of KPR is doubtful for the following reasons.  
First, estimates of various authors depend on data sets and are not in agreement even within the errorbars (see Eq.~\ref{eq-KPR}, \ref{eq-KPR2}, and \ref{eq-KPR3}). 
Second, previous estimates are based on older and thus less complete data sets. 
Today we are able to characterise also many long period objects with low amplitudes, so we simply have better data -- better sampling, better sensitivity, longer baselines.
Third, we believe that the quality of our data is sufficient to rely on this result. It is of course possible that there are some errors in the magnitude level estimates at least from some amateur observations. However, amplitudes are their relative values which are more reliable, since they annul the possible uncertainties in calibration. The remaining errors are estimated at the sub-magnitude level which should not influence the KPR significantly. Another argument confirming the sufficiency of the data quality is the fact that other correlations are present in our data.

To conclude, we challenge the existence of the Kukarkin-Parenago Relation for dwarf novae.
Of course such correlation can be expected, since amplitudes of outbursts are related to the mass of the disk and such mass must be related to the accumulation time. 
Moreover, there is some trend between amplitudes and cycle lengths for these objects for sure, as all the fits presented above give positive slopes and more or less comparable interceptions. However, we question the statement that this trend is a strict correlation.

On the other hand, we found a correlation for superoutbursts amplitudes and their supercycle lengths. It is very interesting because this is in fact a variation of the standard Kukarkin-Parenago Relation which we question. Only this time we consider superoutbursts instead of normal (short) outbursts.

The linear fit to the data of all subtypes of SU~UMa stars (Fig.~\ref{fig-ASO-Psc}) gives:
\begin{equation}
A_{SO} = 2.12(0.22) \cdot \log P_{sc} - 0.45(0.55)
\end{equation}
with $r=0.65$, $R^2=0.42$, and the standard error of the estimate equal to 1.09 which indicate presence of a correlation here. 

\begin{figure}
\centering
\includegraphics[width=0.5\textwidth]{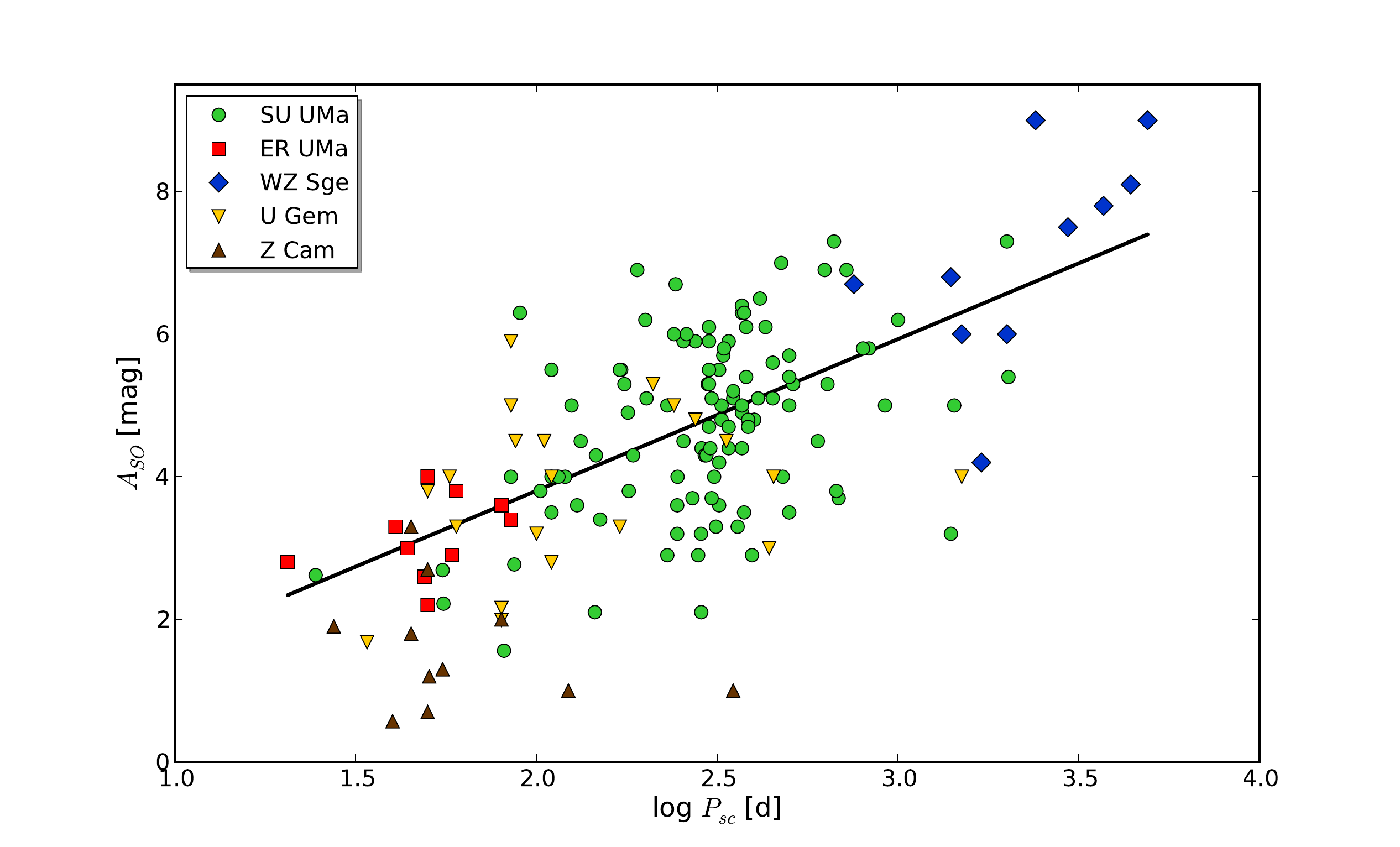}
\caption{log $P_{sc}$ vs. $A_{SO}$}
\label{fig-ASO-Psc}
\end{figure}

A clear distinction between respective subclasses of SU~UMa stars is clearly visible in this plot. ER~UMa stars are concentrated in the lower left part of the plot, WZ~Sge in the upper right side, and typical SU~UMa stars in between. All the green points in the region of ER~UMa stars are the OGLE new objects classified as SU~UMa stars. They most likely require a detailed study to verify this classification. These are particularly interesting stars for future studies. 

Another peculiar feature is the fact that characteristics of long outbursts of long-period systems again seem to be very similar to SU~UMa stars' superoutbursts.  Although this is not a very tight correlation, the trend is surely visible, at least for U~Gem stars.

\subsection{Bailey Relation}
\label{najwazniejsza3}

In this section we discuss the relation between the orbital period and the rate of decline of outbursts ($\tau_d$~[d/mag]) which is constant for each particular object and independent of outburst duration. 
\cite{1975Bailey} was the first to notice this correlation. 

According to \cite{2003Warner}:
\begin{equation}
\log \tau_d [\rm{d~mag^{-1}}] = -0.28 + 0.84 \cdot \log P_{orb} [\rm{h}]
\label{bailey}
\end{equation}

In our catalog we have neither information on decline rates of normal outbursts of SU~UMa-type stars, nor on short outbursts of U~Gem and Z~Cam stars. 
They were not recorded because of a poor quality of most of light curves for normal (short) outbursts. One of our future plans is to collect information on $\tau_d$ even for only a part of our sample, if it is available, to test the Bailey Relation properly.

However, we recorded the needed parameters to find the rate of decline for super- and long outbursts. Figure~\ref{fig-Porb-taud} presents this correlation. 

\begin{figure}
\centering
\includegraphics[width=0.5\textwidth]{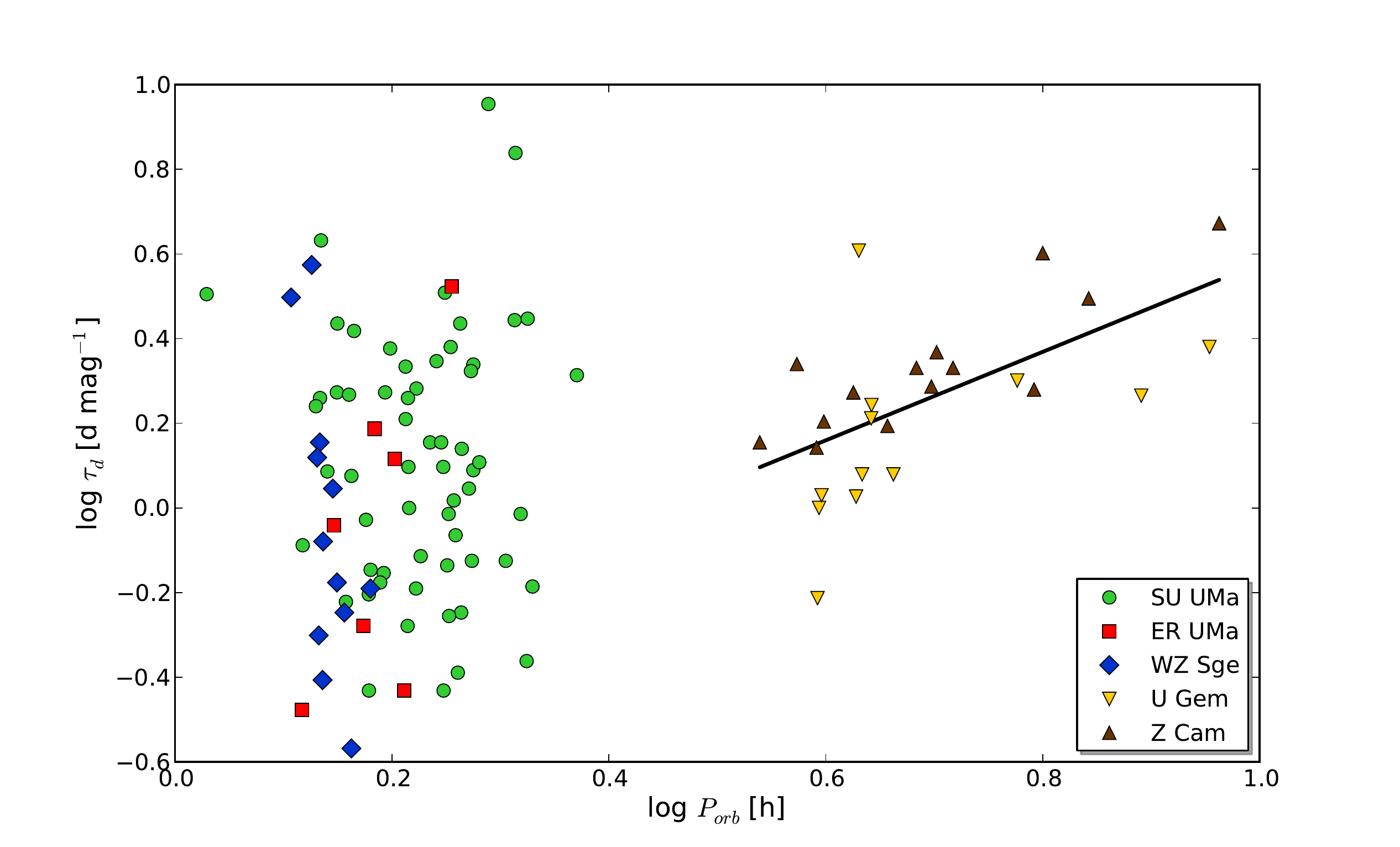}
\caption{ log  $P_{orb}$ vs. log $\tau_d$}
\label{fig-Porb-taud}
\end{figure}

Dwarf novae below and above the period gap create two distinct groups in this plot. 
For short-period stars we found: 
$a = 0.45 \pm 0.55 $,
$b = -0.02 \pm 0.12 $,
$r =  0.09 $,
$R^2 = 0.01$,
and the standard error of the estimate of 0.31 (see the Appendix for the parameters' description). These numbers clearly show that there is no Bailey relation for superoutbursts of SU~UMa stars.

In turn, above the period gap, the situation is different. 
We found:
\begin{equation}
\log \tau_d [\rm{d~mag^{-1}}] = -0.47(0.20) + 1.05(0.28) \cdot \log P_{orb}~[\rm{h}]
\end{equation}
with corresponding coefficients: $r= 0.61$, $R^2=0.37$, and the standard error of the estimate equal to 0.16. Thus, there is a strong correlation for long outbursts of U~Gem and Z~Cam stars. 

Our relation is in agreement within the errorbars with the commonly used equation (Eq.~\ref{bailey}) in spite of the fact that it is derived only based on super- and long outbursts. In the future we plan to improve it with additional data on normal (short) outbursts. 

A similar relation was found for rates of rise of DN outbursts ($\tau_r [\rm{d~mag^{-1}}]$), as reported by \cite{2003Warner}:
\begin{equation}
\log \tau_r [\rm{d~mag^{-1}}]= -0.85 + 1.15 \cdot \log P_{orb} [\rm{h}]
\end{equation}

Here again we have on our hands only data for superoutbursts and long outbursts. We show it in Fig.~\ref{fig-Porb-taur}. 

\begin{figure}
\centering
\includegraphics[width=0.5\textwidth]{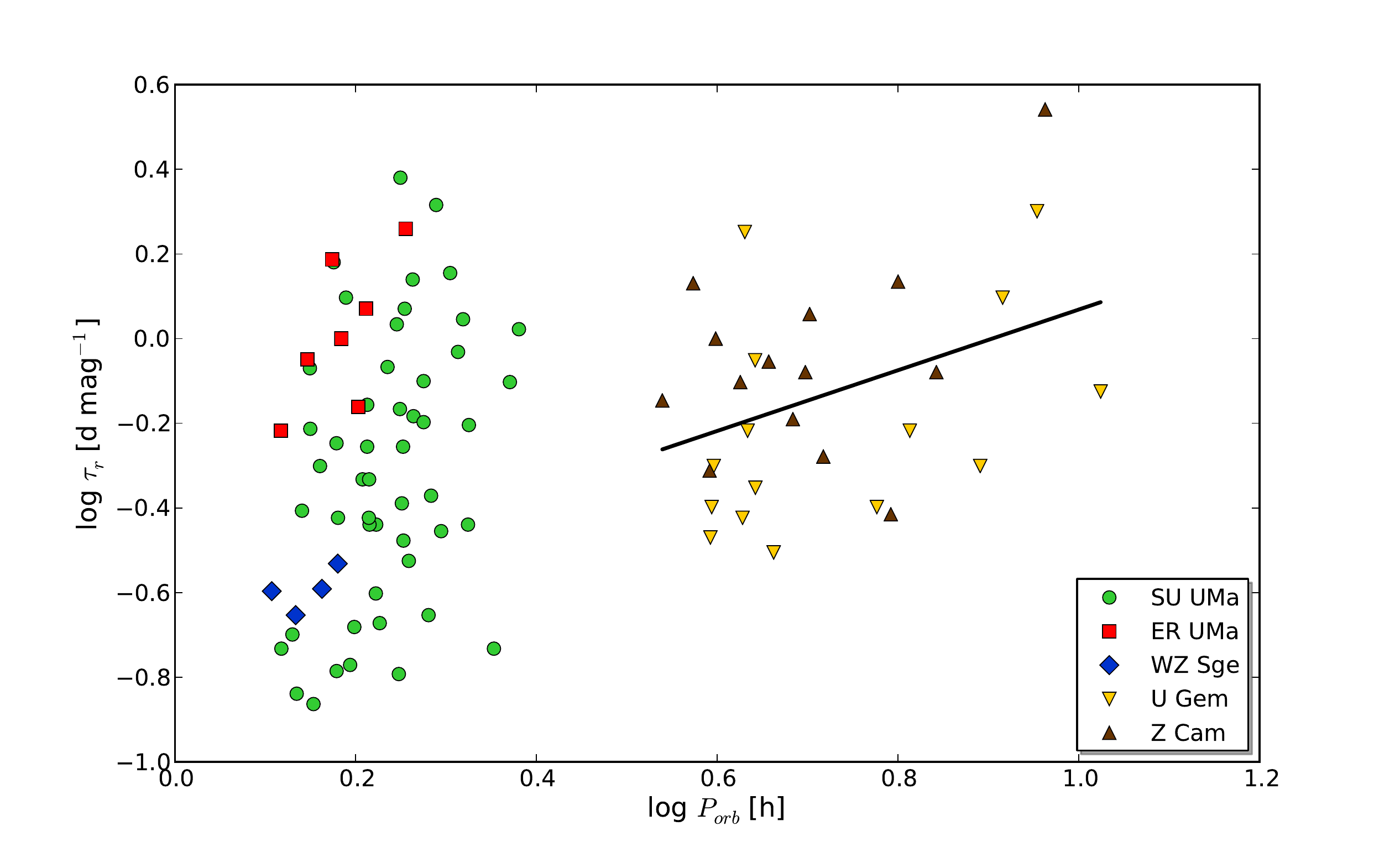}
\caption{log  $P_{orb}$ vs. log $\tau_r$}
\label{fig-Porb-taur}
\end{figure}

Above the period gap we found:
\begin{equation}
\log \tau_r  [\rm{d~mag^{-1}}] = -0.65(0.25) + 0.72(0.34) \cdot \log P_{orb} [\rm{h}]
\end{equation}
with $r= 0.38$, $R^2=0.14$, and the standard error of the estimate of 0.23. This relation is looser, as compared to the standard Bailey relation. 
It may be caused by the fact that $\tau_r$ is shorter and in principal more difficult to measure accurately than $\tau_d$. 
In the future we hope to improve it also with additional measurements for normal (short) outbursts. 

Although there is no clear relation below the gap in the Fig.~\ref{fig-Porb-taur}, there is a tendency between ER~UMa and WZ~Sge stars showing that ER~UMa stars have definitely higher rates of rise of superoutbursts than WZ~Sge stars. 
There is no such visible trend for typical SU~UMa stars, though. 
What makes it even more interesting, this trend is completely not apparent in the previous relation between the rate of decline for super- and long outbursts and the orbital period (Fig.~\ref{fig-Porb-taud}).

\subsection{Other interesting correlations}
\label{exam-corr}

In this subsection we present a few other interesting correlations in our sample.

\subsubsection{Cycle length vs. supercycle length}
\label{PscPc}

The relation between the cycle and supercycle lengths is known since over twenty years. 
\cite{1995Patterson} found: 
\begin{equation}
\log P_{sc} = 1.30 + 0.68~\log P_c
\end{equation}

Our data gives the relation which is in a perfect agreement with the Patterson's finding:
\begin{equation}
\log P_{sc} = 1.32(0.08) + 0.66(0.05)~\log P_c
\end{equation}
This correlation is strong, as the standard error of the estimate is 0.24, the Pearson coefficient $r=0.84$, and $R^2=0.71$, see Fig.~\ref{fig-Pc-Psc}. 

\begin{figure}
\centering
\includegraphics[width=0.5\textwidth]{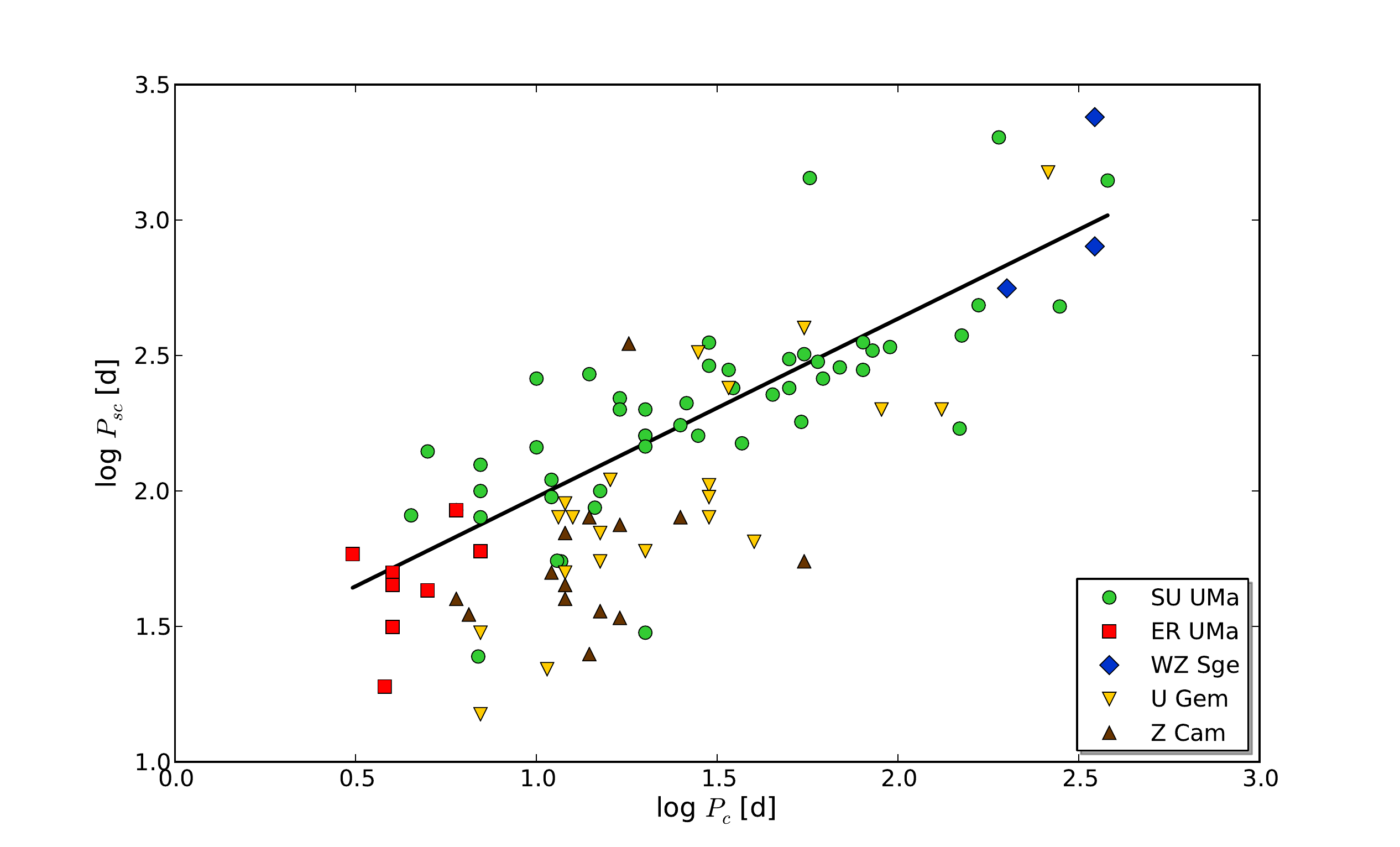}
\caption{log $P_c$ vs. log $P_{sc}$}
\label{fig-Pc-Psc}
\end{figure}

Like everywhere else in our work, when it comes to parameters typical for superoutbursts  (e.g. $P_{sc}$) we use only data for SU~UMa stars, without U~Gem and Z~Cam-type objects, to find the fit to the data. 
This is because long outbursts of U~Gem and Z~Cam stars fundamentally differ from superoutbursts of SU~UMa stars, according to the TTI model. However, Fig.~\ref{fig-Pc-Psc} shows that this relation is most likely true also for these dwarf novae above the period gap. We can at least say that they do not deviate from the fit significantly. This is somehow surprising, since long and super-outbursts are so different in other ways.

\subsubsection{Amplitudes of outbursts}
\label{ANO-ASO}

\begin{figure}
\centering
\includegraphics[width=0.5\textwidth]{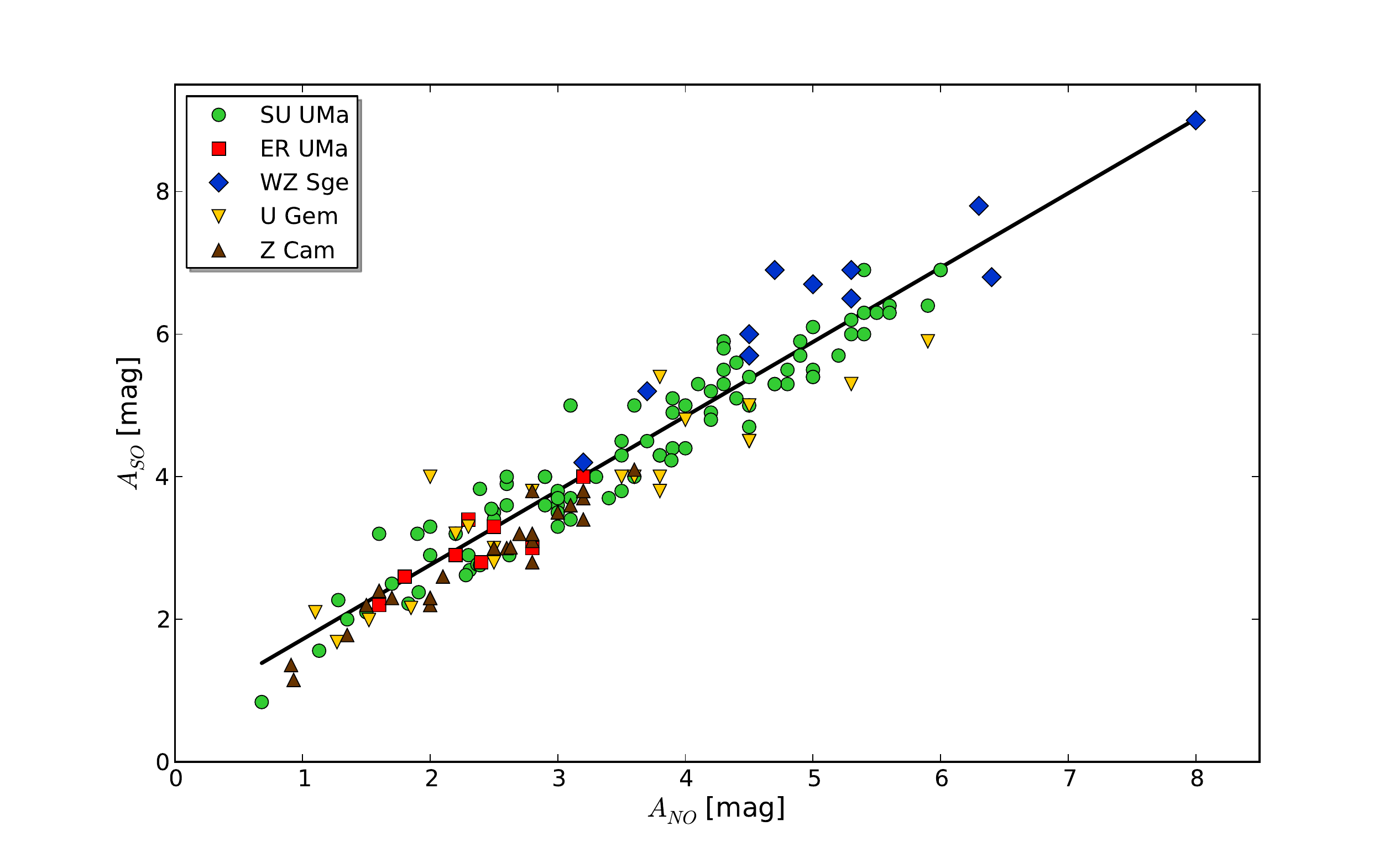}
\caption{$A_{NO}$ vs. $A_{SO}$}
\label{fig-ANO-ASO}
\end{figure}

\begin{figure}
\centering
\includegraphics[width=0.5\textwidth]{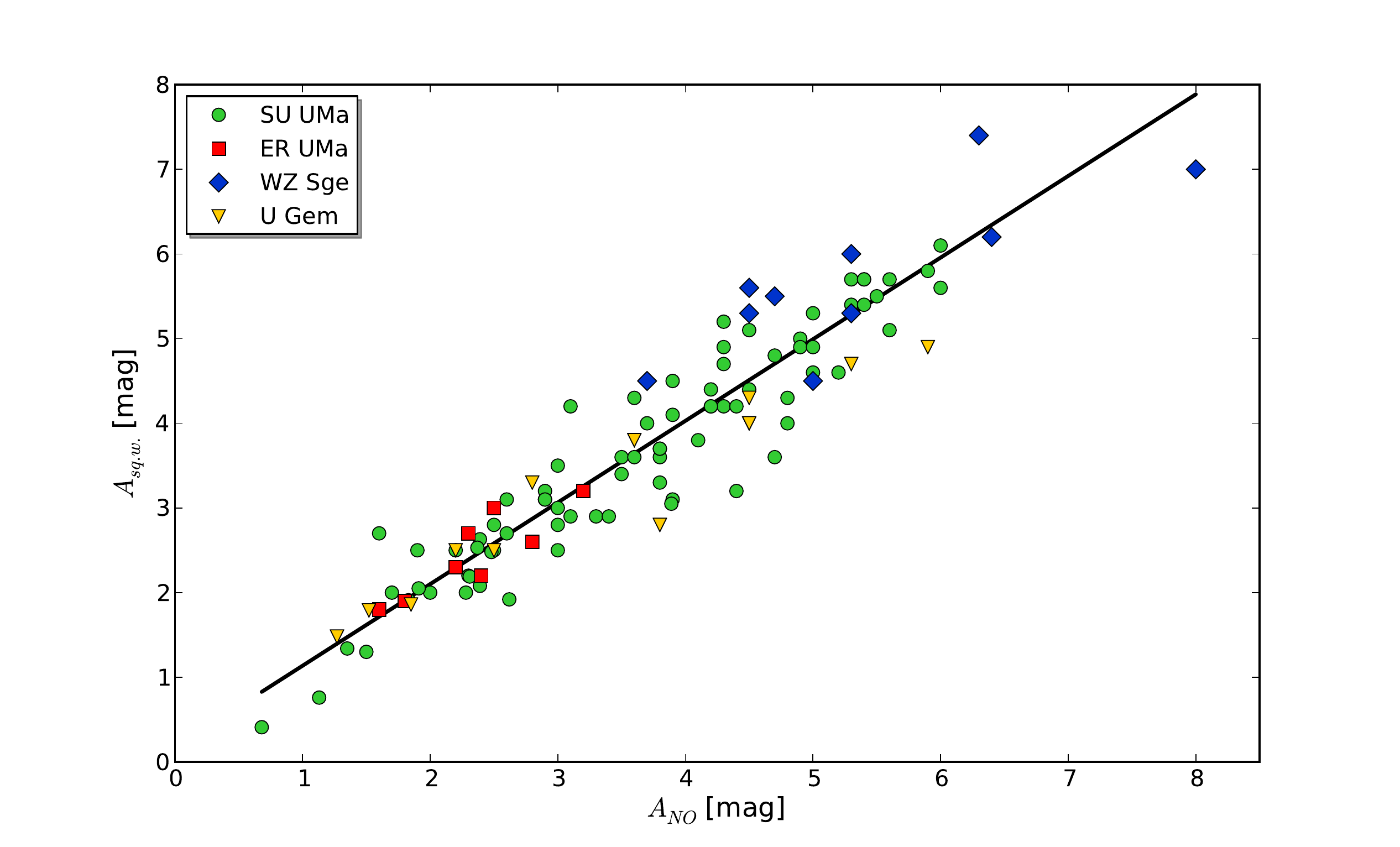}
\caption{$A_{NO}$ vs. $A_{sq.w.}$}
\label{fig-ANO-Asqw}
\end{figure}

Figures~\ref{fig-ANO-ASO} and \ref{fig-ANO-Asqw} show a clear relation between amplitudes of normal (or short) outbursts and amplitudes of two representatives of super- (or long) outbursts. 
Formal fits to the data points give the relations: 
\begin{equation} A_{SO} = 1.043(0.029)\cdot A_{NO} + 0.68(0.11)	\end{equation}
\begin{equation} A_{sq.w.} = 0.964(0.034)\cdot A_{NO} + 0.17(0.13)  \end{equation} 
The standard errors of both estimates are 0.41 and 0.47, respectively.

The first one reveals $r$ and $R^2$ of 0.96 and 0.92 and the latter one of 0.94 and 0.89,  respectively. 
Thus, for both of them the hypothesis that they are unrelated can be easily rejected at 0.05 level of significance. Dependence on $A_{SO}$  gives a more precise equation, whereas the second one shows that $A_{NO}$ and $A_{sq.w.}$ are almost exactly equivalent. 

It is worth mentioning that these relations, although derived only based on SU~UMa stars data, are true for both amplitudes of superoutbursts of SU~UMa stars as well as of long outbursts of U~Gem and Z~Cam objects. They are all shown in both figures with the exception of Z~Cam stars for the $A_{NO}$ vs. $A_{sq.w.}$ relation because there is no record of their $A_{sq.w.}$ (see Sect.~\ref{phot-feat}). 

To conclude, when it comes to amplitudes of outbursts, all types of dwarf novae behave alike.

\subsubsection{Amplitude vs. outburst duration}
\label{ANO-DNO}

The $A_{NO}$ vs. $D_{NO}$ relation (Fig.~\ref{fig-ANO-DNO}) gives the formal fit for data of SU~UMa stars:
\begin{equation} D_{NO} = 0.06(0.14)\cdot A_{NO} + 3.7(0.5)  \end{equation}
with the standard error of the estimate equal to $1.42$, $r=0.057$, and $R^2=0.003$, which means that the null hypothesis in this case cannot be rejected at the given level of significance. 
Thus, this \textit{correlation is not true}. There is obviously no correlation for U~Gem and Z~Cam stars as well. 

\begin{figure}
\centering
\includegraphics[width=0.5\textwidth]{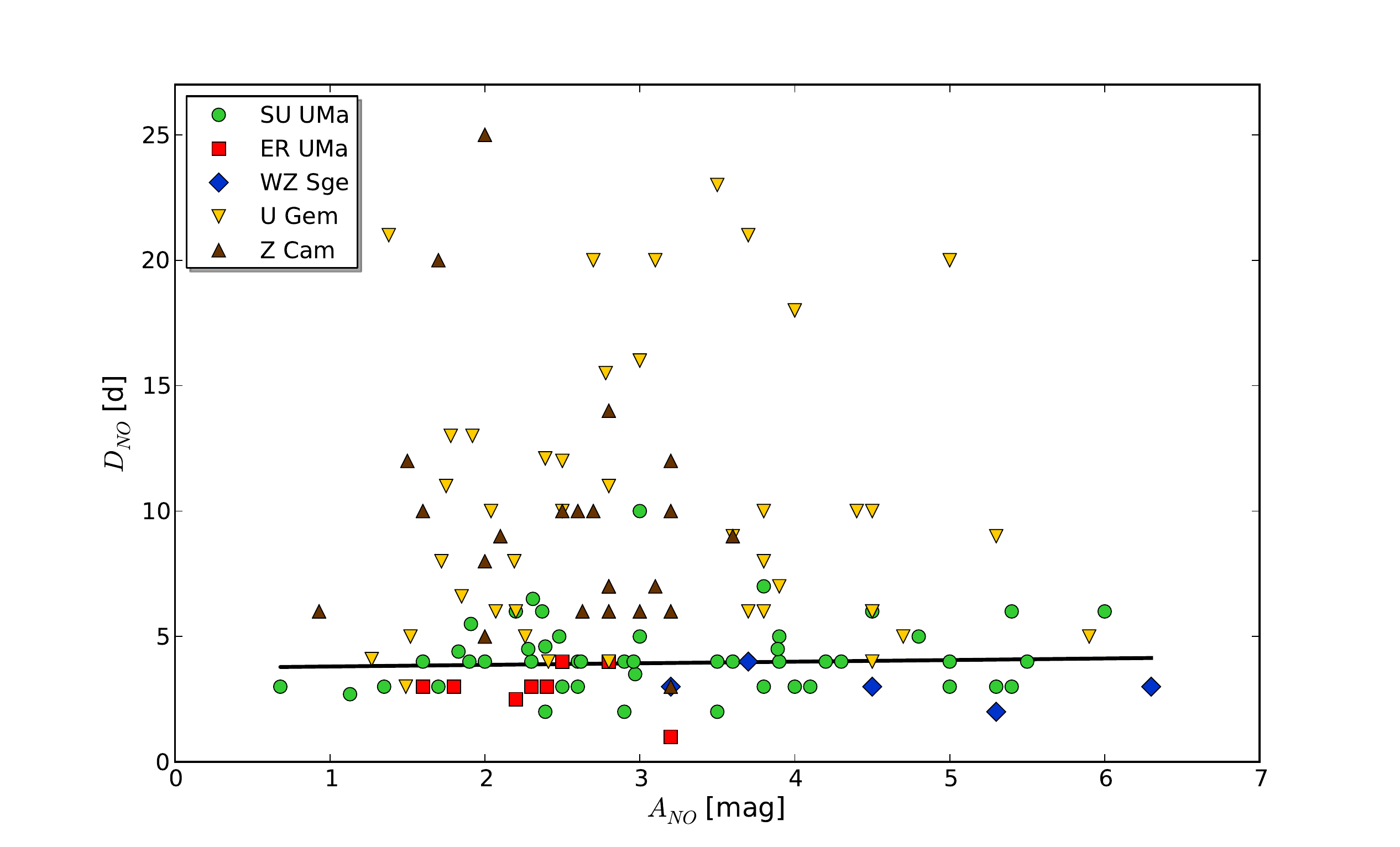}
\caption{$A_{NO}$ vs. $D_{NO}$}
\label{fig-ANO-DNO}
\end{figure}

\begin{figure}
\centering
\includegraphics[width=0.5\textwidth]{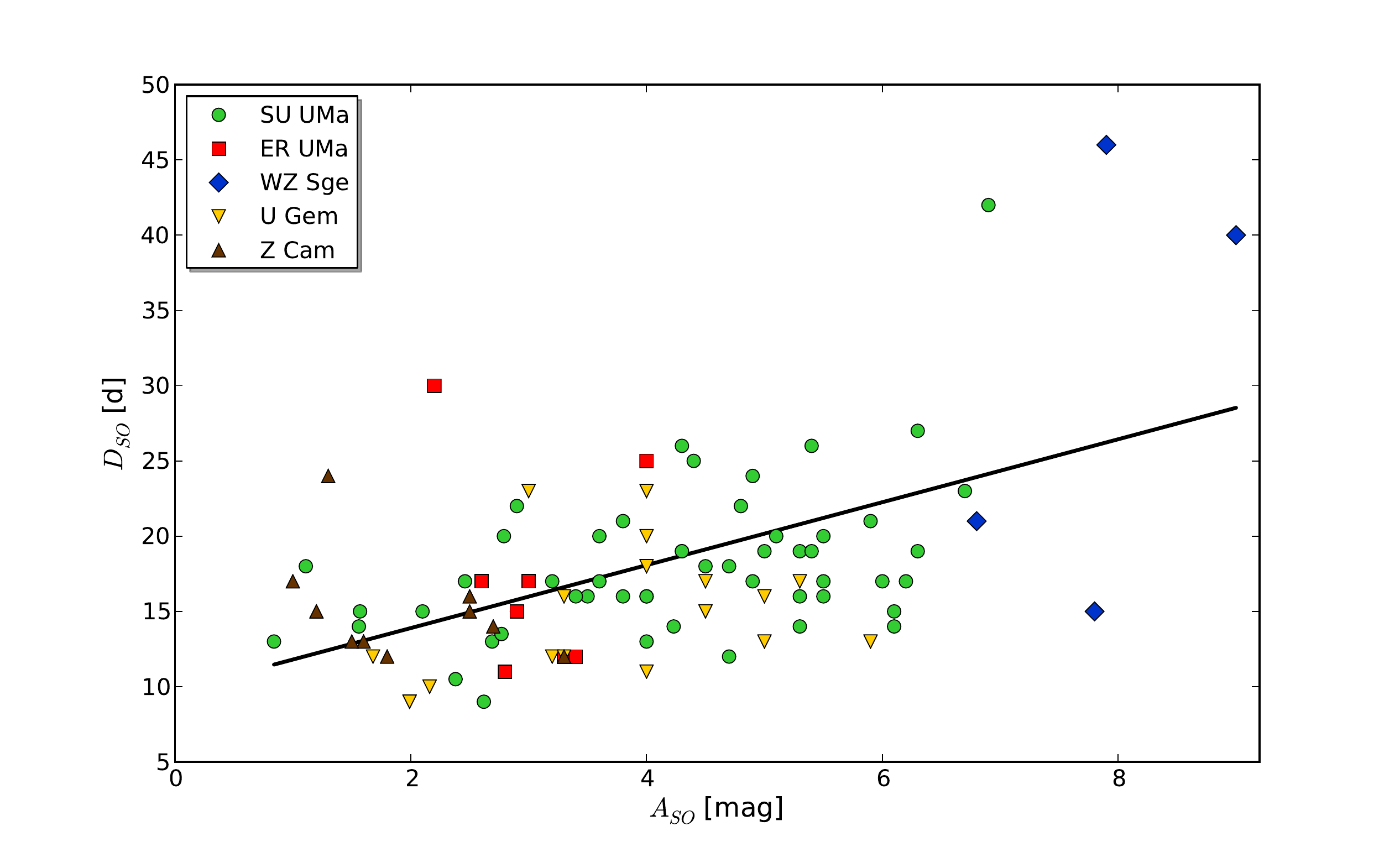}
\caption{$A_{SO}$ vs. $D_{SO}$}
\label{fig-ASO-DSO}
\end{figure}

The scatter plot shows that SU~UMa stars' normal outbursts duration is rather fixed at a value mostly close to $\sim~4$~days, whereas for long-period objects this value is random and span over the range of about 20 days. Thus, in the context of normal (short) outbursts duration there is a clear discrepancy between DN above and below the period gap, unlike in the case of their amplitudes. 

It is possible that the scatter of  long-period objects may be partially the effect of a wrong classification between short and long outbursts of U~Gem and Z~Cam stars. However, it is not likely, since we were very careful in this matter during the measurements. Another explanation of this scatter could be the fact that the measured values of $D_{NO}$ are typical for each star but they are not always constant. Thus, this effect may be partially also a manifestation of the variability of short outbursts duration for stars above the period gap.

When it comes to superoutbursts (or long outbursts in the case of U~Gem and Z~Cam stars), their amplitudes and durations seem to be related, see Fig.~\ref{fig-ASO-DSO}. 

We found the fit to data of SU~UMa stars:
\begin{equation}
D_{SO} = 2.09(0.43)\cdot A_{SO} + 9.7(2.0)  
\end{equation}
with $r=0.52$, $R^2=0.27$, and the standard error of the estimate of 5.7, which stand for a true correlation with the 0.05 level of significance.
Here again, the long-period systems behave similar to SU~UMa-type stars, since they apparently follow the given relation. 

In this context, a comparison of Figures~\ref{fig-ANO-DNO} and \ref{fig-ASO-DSO} is interesting because they strongly differ from each other for U~Gem and Z~Cam stars. 
In the case of short outbursts there is no correlation with amplitudes, whereas long outbursts are correlated with amplitudes. Thus, both these types of outbursts of U~Gem and Z~Cam stars should be treated separately in this kind of analysis.

\subsubsection{Outburst duration vs. orbital period}

\cite{2000Smak} presented the relation between widths of outbursts versus the orbital period for narrow (normal or short) as well as wide (super- and long) outbursts. He listed the explanation of this relation as one of the most important unsolved problems of dwarf nova outbursts. In Figures~\ref{fig-Porb-DNO} and \ref{fig-Porb-DSO} we plot the same relations based on our data and thus confirm the conclusions of \cite{2000Smak}.

\begin{figure}
\centering
\includegraphics[width=0.5\textwidth]{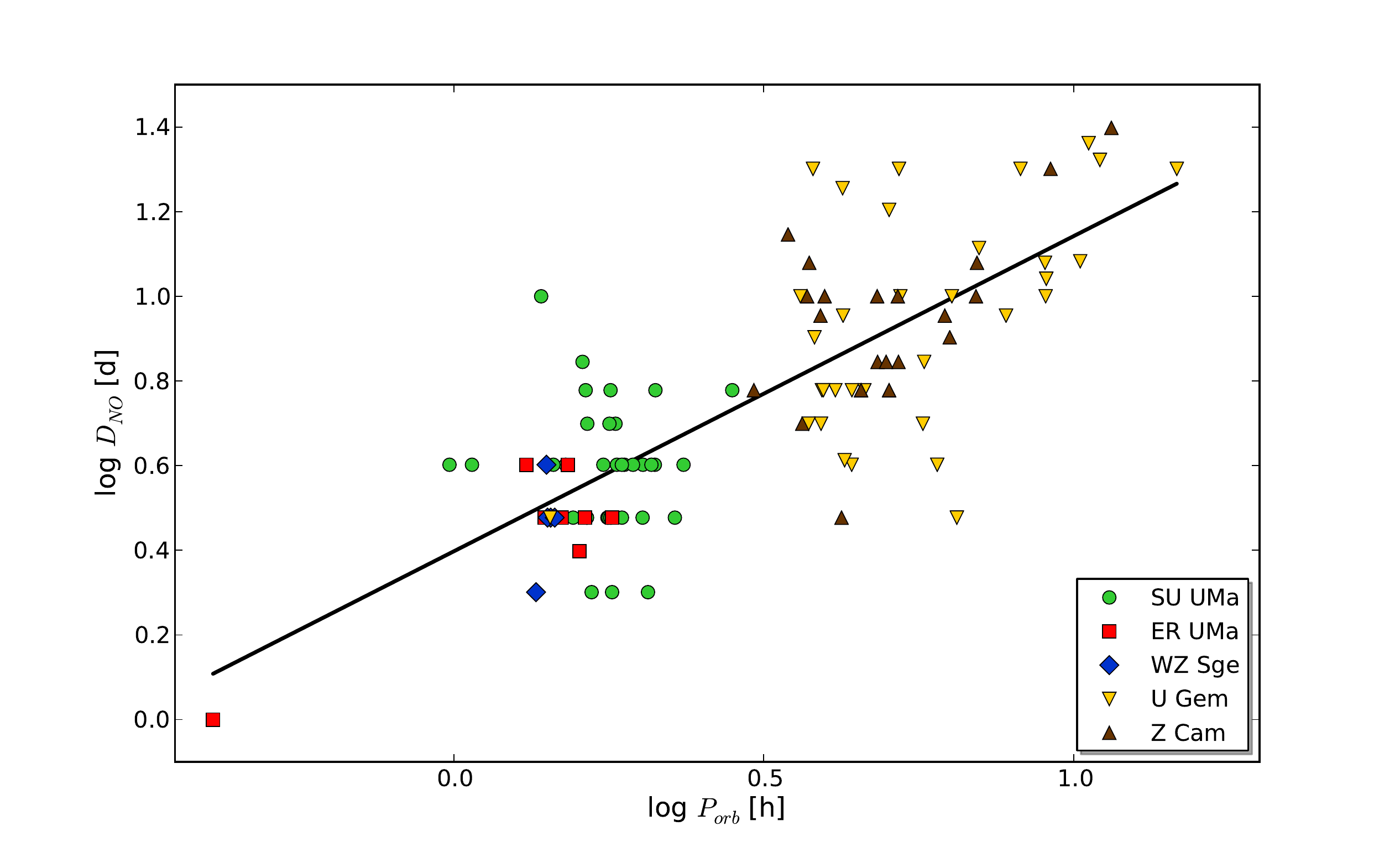}
\caption{log $P_{orb}$ vs. log $D_{NO}$}
\label{fig-Porb-DNO}
\end{figure}

\begin{figure}
\centering
\includegraphics[width=0.5\textwidth]{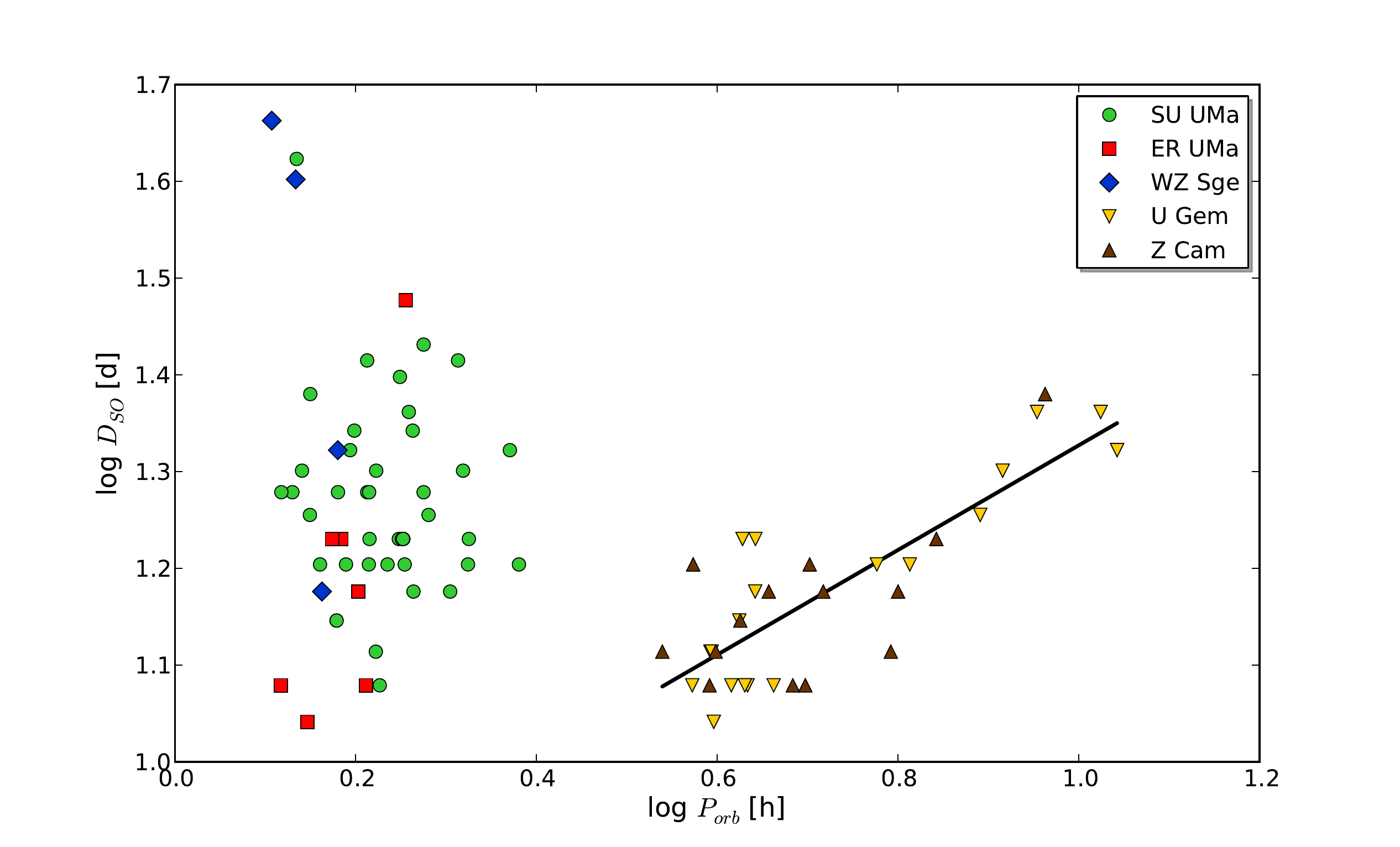}
\caption{log $P_{orb}$ vs. log $D_{SO}$}
\label{fig-Porb-DSO}
\end{figure}

For the relation between $P_{orb}$ and $D_{NO}$ we found: 
\begin{equation}
\log D_{NO} = 0.745(0.060) \cdot \log P_{orb} + 0.398(0.033)
\end{equation}
with $r=0.78$, $R^2=0.61$, and the standard error of the estimate of 0.18, which confirm the existence of the relation.

In turn for $P_{orb}$ vs. $D_{SO}$ we found the fit for dwarf novae above the period gap:
\begin{equation}
\log D_{SO} = 0.541(0.068) \cdot \log P_{orb} + 0.786(0.049)
\end{equation}
with $r=0.82$, $R^2=0.67$, and the standard error of the estimate of 0.05, which also confirm the existence of the relation.

Both these relations are in agreement with Fig.~1 of \cite{2000Smak}. With a larger data set we confirm that the correlation for narrow outbursts is true for both short- and long-period objects. 
We also confirm that durations of superoutburst do not follow the relation of long outbursts of U~Gem and Z~Cam stars as a function of orbital period. 
What is more, there seems to be no correlation for outbursts durations of SU~UMa stars in this domain at all.

\subsubsection{Outburst duration vs. mass ratio}

Another interesting correlation is found between duration of normal (short) outbursts and the mass ratio (Fig.~\ref{fig-DNO-q}). 
It follows from the correlation between outburst duration and orbital period, since secondary mass varies with period in CVs.

\begin{equation}
\log D_{NO} = 0.684(0.069) \cdot \log q + 1.152(0.053)
\end{equation}
with $r=0.78$, $R^2=0.61$, and the standard error of the estimate of 0.18, which also confirm the existence of the relation.

\begin{figure}
\includegraphics[width=0.5\textwidth]{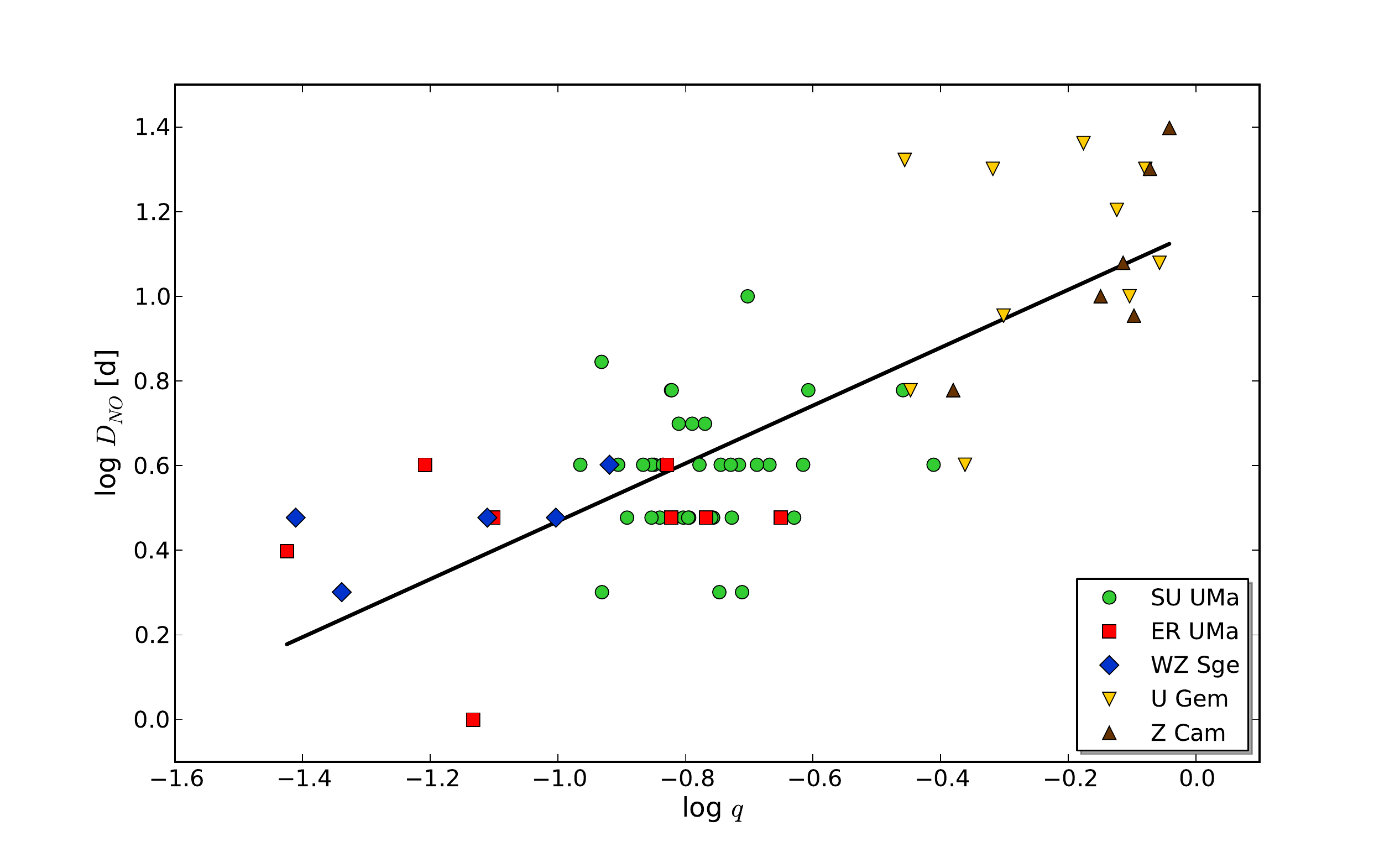}
\caption{log $q$ vs. log $D_{NO}$}
\label{fig-DNO-q}
\end{figure} 

When it comes to the same relation for superoutbursts, with the obtained Pearson parameter we cannot reject the null hypothesis that there is no correlation in this case. See Fig.~\ref{figqDSO}. 

\begin{figure}
\includegraphics[width=0.5\textwidth]{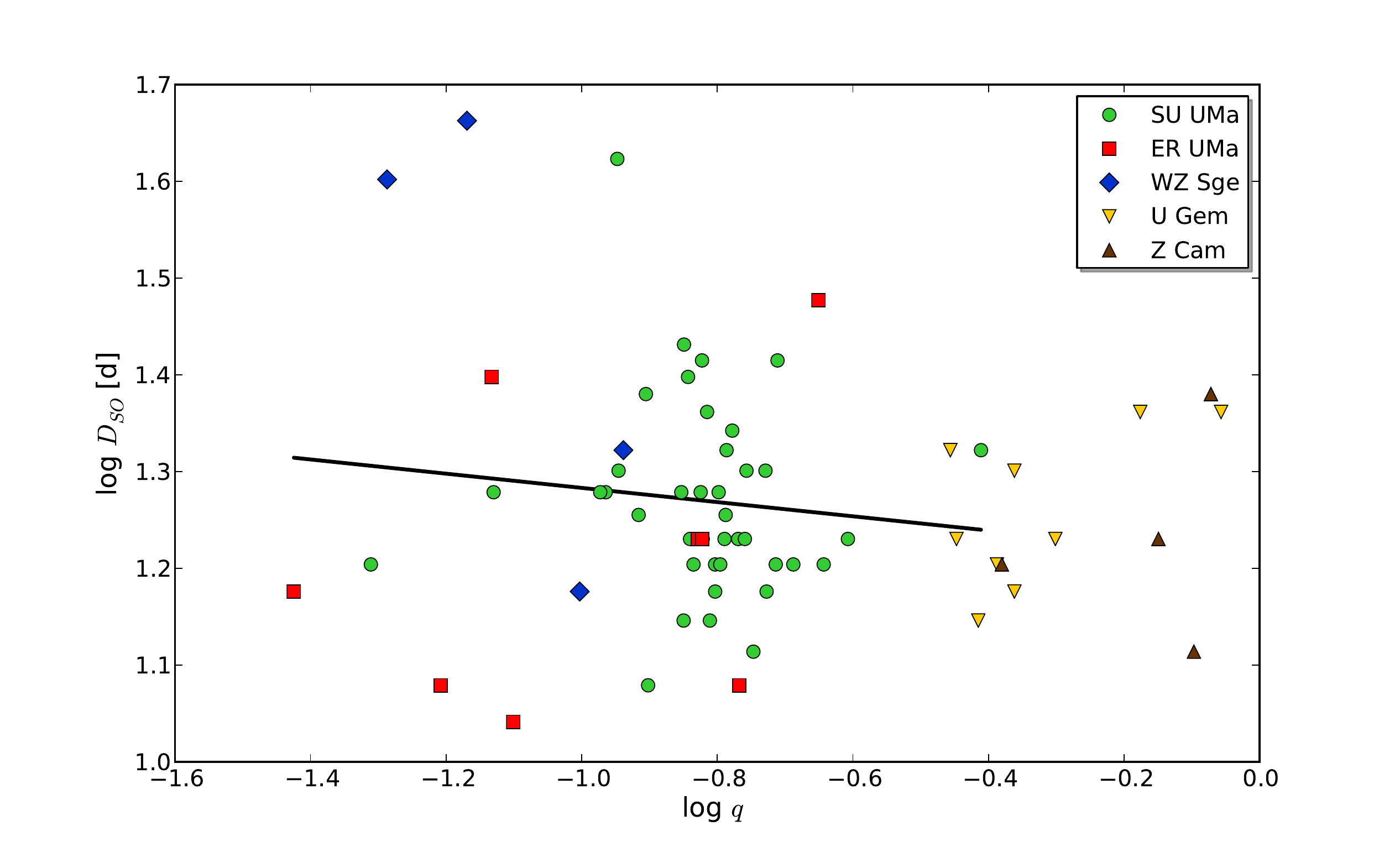}
\caption{log $q$ vs. log $D_{SO}$}
\label{figqDSO}
\end{figure}

This is surprising, especially in the context of the TTI model.
After the discovery of the first objects of WZ~Sge type, which are characterized by a low mass ratio and the longest superoutburst durations, i.e., located in the upper left region of Fig.~\ref{figqDSO}, 
we would expect that objects with shorter outbursts will have a larger mass ratio and will occupy opposite regions of the plot.

Normal outbursts are understood as caused by the thermal instability alone, while during superoutburst additionally the tidal instability occur, according to the TTI model of \cite{1996Osaki}. 
In this theory the critical resonance radius, needed for the tidal instability 
to take place, occurs only for systems with the mass ratio $q<0.25$. 
\cite{2005Osaki} 
pointed that conditions which determine which type of superoutburst occur are the growth rate of an eccentric disk by the tidal instability and the rate of shrinkage of the disk below the 3:1 resonance after a normal outburst. They both depend on the mass ratio. Thus, we would expect at least that individual subtypes of superoutburst (that means also subclasses of SU~UMa-type stars) will be concentrated in similar regions of the plot. 
This is not what we observe. Theoretical models should tackle this problem.

\subsubsection{$\tau_{r}$ vs. $\tau_{d}$}
\label{tau_r-tau_d}

The linear fit for $\tau_{r}$ vs. $\tau_{d}$ gives:
\begin{equation}
\tau_{d} = 1.02(0.19) \cdot \tau_{r} + 0.83(0.19)
\label{eq-tau}
\end{equation}
with $r=0.57$, $R^2=0.33$, and the standard error of the estimate of 0.94, which also confirm the existence of the relation (Fig.~\ref{fig-tau_r-tau_d}).

\begin{figure}
\centering
\includegraphics[width=0.5\textwidth]{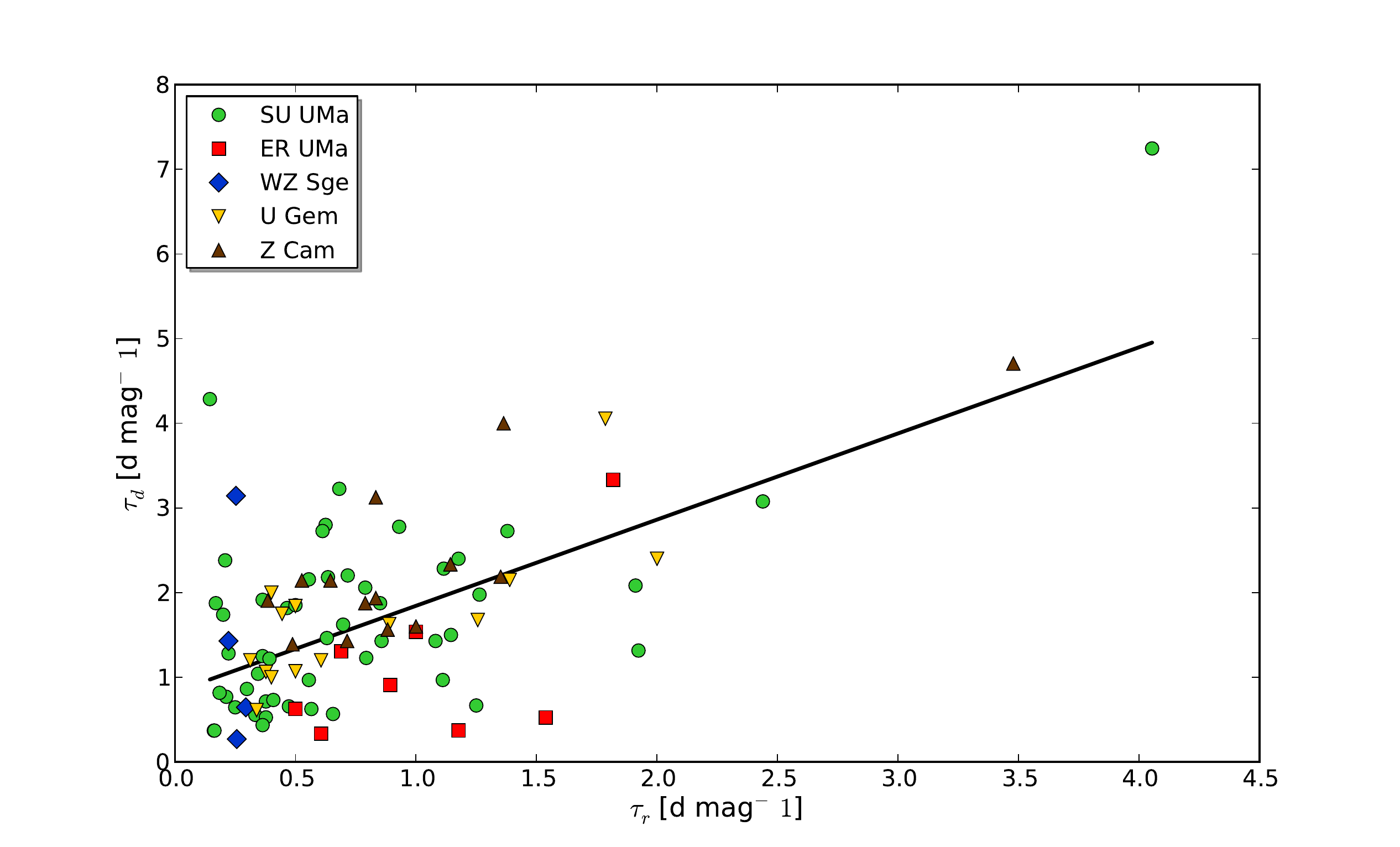}
\caption{$\tau_r$ vs. $\tau_{d}$}
\label{fig-tau_r-tau_d}
\end{figure}

\cite{2003Warner} reported a linear correlation between $\tau_r$ and $\tau_{d}$, where  $\tau_r\sim\frac{1}{3}\tau_{d}$ for $P_{orb}<2.1$~h, and 
$\tau_r\sim\frac{1}{2}\tau_{d}$ for $P_{orb}>3$~h.
DN in our sample seem to follow the fit (Eq.~\ref{eq-tau}) regardless of the orbital period. Here again, the fit was obtained only from SU~UMa data but long-period stars agree with the fit nicely. 
However, the scatter of this relation is quite significant and we cannot exclude such a distinction for dwarf novae from each side of the period gap. 

In principle, the longer the rise of superoutburst, the longer the decline. 
$\tau_{d}$ take on average about one day more to change the brightness by one magnitude, as compared to $\tau_{r}$.

\section{Conclusions}
\label{sec-relations-con}

The first and main goal of this work was to measure and collect all the accessible photometric features of outburst and superoutbursts of dwarf novae. 
Then, we also performed the first approach to study the created catalog. 
We searched for correlations between all possible observables and physical parameters of such systems, based on their light curves. 
Our aim was to check the known relations between some of them and to look for potential new relations between others. We analyzed as many properties of DN as possible, such as their orbital period, cycle's and supercycle's lengths, amplitudes and duration of normal and superoutbursts, mass ratio, rate of brightness rise and decrease in superoutbursts.

The main findings of this part of work are the following.
\begin{itemize}
\item Confirmation of several known relations, for instance the dependence of normal cycle length on supercycle length.
\item Contesting the existence of the Kukarkin-Parenago relation. However, we found its equivalent in the form of $P_{sc}$ vs. $A_{SO}$ relation. 
\item Finding no presumed relation in the case of the $q$~vs.~$D_{SO}$. 
\end{itemize}

We found no correlation between inclination with any other parameter in our study. 
Correlations with the superhump period are also not shown because they strongly resemble these of the orbital period. 
What is more, we cannot also draw any conclusion based on the occurrence or absence of precursors of superoutbursts, since we were able to notice only a few of them because most superoutbursts observations are carried on only after the maximum brightness. 

We plan to continue a detailed work on the correlations listed in the Appendix. We find this goal very interesting and needed, since the number of cataclysmic stars is growing exponentially over the past years. 
The correlations which are being used in the community for decades now are supposed to be confronted with the newest, often more precise, data.

We hope that this work will help to direct theoretical studies and decrease discrepancy between observations and theory. 

\section*{Acknowledgements}

This project was supported by Polish National Science Center grant DEC-2011/03/N/ST9/03289.
We are very grateful to an anonymous referee for the thoughtful comments which helped to significantly improve the text of the paper.




\bibliographystyle{mnras}
\bibliography{bibliografia} 



\appendix

\section{Other correlations}

Here we present formal linear fits to other correlations which were found in our data. 
Each relation is given in the form of a linear function: 
\begin{equation}
y(x) = a(\pm SE_a) \cdot x + b(\pm SE_b)
\end{equation}
where $SE_a$ and $SE_b$ are the standard errors of the parameters $a$ and $b$.
For each relation the coefficients and variables are listed in the Table together with other values of the parameters, namely:
the standard error of the estimate, i.e., $SE_e$, $r$, $R^2$, and a Yes[Y] or No[N] information indicating if the Pearson coefficient fulfilled the conditions that the null hypothesis can be rejected at the given level of significance ($\alpha = 0.05$), which means that there is a correlation in the statistical sense.

\begin{table*}
\caption{List of parameters of linear fits and statistical information for all correlations in our sample. (1) stands for a fit for SU~UMa-type stars, (2) indicates a fit for other types of dwarf novae. \vspace{0.5cm}}
\begin{tabular}{@{}l|l|c|c|c|c|c|c|c|c@{}}
$x$ 		& 	$y$ 			&	$a$	& $SE_a$	& $b$	& $SE_b$	&	$SE_e$ & $r$	&	 $R^2$	&	$H_0$  \\
	& 			&	&	& 	&	&	 & 	&		&rejected \\
\hline
$A_{NO}$	 &	$D_{P}$		& 1.34 	& 0.28 		& 4.8   & 1.0			& 3.66 		& 0.44 	& 0.19 	& Y \\
$A_{SO}$	 &	$D_{P}$		& 1.42 	& 0.19 		& 4.10 	& 0.86 			& 3.40 		& 0.56 	& 0.31 	& Y \\
$A_{sq.w.}$&	$D_{P}$		& 1.42 	& 0.26 		& 4.8 	& 1.0 			& 3.83 		& 0.45 	& 0.20 	& Y \\
$A_{NO}$	 &	$D_{SO}$		& 2.18 	& 0.54 		& 10.6 	& 2.1 			& 5.62 		& 0.51 	& 0.26 	& Y \\
$A_{sq.w.}$&	$D_{SO}$		& 2.11 	& 0.49 		& 11.2 	& 1.9 			& 5.95 		& 0.49 	& 0.24 	& Y \\
$A_{NO}$	 &$D_{sq.w.}$	& 1.14 	& 0.25 		& 8.39 	& 0.98 			& 3.43 		& 0.42 	& 0.18 	& Y \\
$A_{SO}$	 &$D_{sq.w.}$	& 1.34 	& 0.16 		& 6.90 	& 0.85 			& 3.64 		& 0.53 	& 0.28 	& Y \\
$A_{sq.w.}$&$D_{sq.w.}$	& 1.29& 0.18 		& 8.01 	& 0.78 			& 3.59 		& 0.48 	& 0.23 	& Y \\

log $A_{NO}$	 &	log $\tau_r$		& -1.12 	& 0.13 		& 0.360 	& 0.071 			& 0.19 		& -0.76 	& 0.57 	& Y \\
log $A_{SO}$	 &log 	$\tau_r$		& -1.45 	& 0.13 		& 0.64 	& 0.08 			& 0.21 		& -0.80 	& 0.64 	& Y \\
log $A_{sq.w.}$ &log $\tau_r$	& -1.14	& 0.11 		& 0.333 	& 0.063 			& 0.22 		& -0.77 	& 0.60 	& Y \\
log $A_{NO}$	 &	log $\tau_d$		& -0.79 	& 0.20 		& 0.50 	& 0.11 			& 0.30 		& -0.43 	& 0.19 	& Y \\
log $A_{SO}$	 &	log $\tau_d$		& -0.85 	& 0.14 		& 0.639 	& 0.094 			& 0.30 		& -0.52 	& 0.27 	& Y \\
log $A_{sq.w.}$ &log $\tau_d$	& -0.77	& 0.12 		& 0.516 	& 0.072 			& 0.29 		& -0.54 	& 0.29 	& Y \\
$A_{NO}$	 &log $P_{sc}$	& 0.173 	& 0.031 		& 1.74 	& 0.13 			& 0.39 		& 0.52 	& 0.27 	& Y \\
$A_{sq.w.}$ &log$P_{sc}$	& 0.206 	& 0.023 		& 1.58 	& 0.10 			& 0.33 		& 0.64 	& 0.41 	& Y \\

$D_{NO}$	 &	$D_{P}$ (1)	& -1.00 	&  0.38		& 14.2 	&  1.6			& 3.17 		& -0.37 	& 0.14 	& Y \\
$D_{NO}$	 &	$D_{P}$	(2)	& 0.233 	&  0.076		& 4.45 	&  0.75			& 2.03 		&  0.48	& 0.23 	& Y \\
$D_{NO}$	 &	$D_{SO}$	 (1)	& -0.58 	&  0.64		& 19.4	&  2.6			& 4.79		& -0.15 	& 0.02 	& Y \\
$D_{NO}$	 &	$D_{SO}$	 (2)	&  0.576	&  0.097		& 9.80 	&  0.97			& 2.51		& 0.75	& 0.56 	& Y \\
$D_{NO}$	 &	$D_{sq.w.}$ (1)	& -0.84 	& 0.45 		& 15.6 	&  1.8			&  4.06		& -0.25 	& 0.06 	& N \\
$D_{NO}$	 &	$D_{sq.w.}$	(2)& 0.92  	&  0.19		& 4.3 	&  1.8			&  4.93		& 0.63 	& 0.40 	& Y \\
$D_{D}$	 &	$D_{SO}$		& 1.32  	&  0.11		& 11.73 	&  0.75			&  3.73		& 0.83 	& 0.69 	& Y \\

log $D_{NO}$	 &log 	$P_{c}$		& 0.78   	&  	0.15	&   0.69	&   0.12		&   0.41		& 0.45  	&  0.20 	& Y \\
log $D_{SO}$	 &log 	$P_{c}$		& 1.85   	&  	0.58	&  -1.08 	&  0.73		&   0.49		&  0.45 	&   0.21	& Y \\
log $D_{SO}$	 &log 	$P_{sc}$		&  1.61  	&  0.45		&  0.24 	&   0.56			&   0.40		&  0.44 	&  0.19 	& Y \\

$D_{R}$	 &	$\tau_r$		& 0.389 	&  0.041		& -0.22 	&  0.12			&  0.47		& 0.73 	& 0.54 	& Y \\
log $D_{P}$  &	log $\tau_r$		& -0.99  	& 0.22 		& 0.75 	&  0.22		&  0.29		& -0.48 	& 0.23 	& Y \\
$D_{P}$	 &	$\tau_d$		& 0.031  	& 0.033 		& 1.25 	&  0.39		& 1.21 		& 0.11	& 0.01 	& N \\
$D_{D}$	 &	$\tau_d$		& 0.208  	&  0.051		& 0.76 	&  0.37		&  2.26		& 0.39 	& 0.15 	& Y \\
log $D_{P}$	 &	log $P_{sc}$	& 1.31  		&  0.31		& 0.98 	&  0.33		&  0.38		& 0.44 	& 0.19 	& Y \\
log $D_{sq.w.}$ &log $P_{sc}$& 1.55  	&  0.33		& 0.71 	&  0.37		&  0.40		& 0.40 	& 0.16 	& Y \\
log $\tau_r$ &	log $P_{sc}$	& -0.68  	&  0.15		& 2.109 	&  0.063		& 0.38 		& -0.49 	& 0.24 	& Y \\

log $q$	& $A_{NO}$	  	& -1.42 	& 0.37 		& 2.67 	& 0.28 			& 1.20 		& -0.33 	& 0.11 	& Y \\
log $q$	& $A_{SO}$	 	& -2.34 	& 0.43 		& 3.03 	& 0.39 			& 1.42 		& -0.33 	& 0.11 	& Y \\
log $q$	& $A_{sq.w.}$	& -1.84 	& 0.52 		& 2.75 	& 0.48 			& 1.28 		& -0.27 	& 0.07 	& Y \\
log $q$	 & log $D_{P}$		& -0.225 &  0.084	& 0.848 	&  	0.074		&  0.12		& -0.30 	& 0.09 	& Y \\
log $q$ 	 & log $D_{sq.w.}$	& -0.162  	& 0.045 		& 0.974 	&  0.042		&  0.11		& -0.28	& 0.08 	& Y \\
log $q$ 	 &	log $P_{c}$		& 0.11  	&  	0.17	&  1.50			& 0.13 		& 0.55 	& 0.07 & 0.01	& N \\
log $q$ 	 &	log $P_{sc}$		& -0.62  	&  0.21		& 1.93 	&  	0.18		&  0.41		& -0.26 	& 0.07 	& Y \\

log $P_{orb}$& $A_{NO}$	 & -1.39	& 0.28 		& 4.18 	& 0.16 			& 1.14 		& -0.34 	& 0.12 	& Y \\
log $P_{orb}$& $A_{SO}$	 & -2.72	& 1.01 		& 5.68 	& 0.23 			& 1.52 		& -0.17 	& 0.03 	& Y \\
log $P_{orb}$& $A_{sq.w.}$ & -4.18& 1.46 		& 5.31 	& 0.33 			& 1.31 		& -0.22 	& 0.05 	& Y \\

log $P_{orb}$& $D_{R}$ (1)		& 1.30  	& 0.63 		&  0.4	&  1.1		& 1.26 		& 0.26 	&  0.07	& Y \\
log $P_{orb}$& $D_{R}$ (2)		& 0.34  	& 0.13 		& 0.96 	&  0.79		& 1.69 		& 0.41 	& 0.17 	& N \\

log $P_{orb}$& $D_{P}$ (1)		& -3.7  	&  1.6		& 17.7 	&  2.7		&  3.71		& -0.26 	& 0.07 	& Y \\
log $P_{orb}$& $D_{P}$ (2)		& 0.61  	&  0.17		& 3.6 	&  1.0		&  1.96		& 0.52 	& 0.27 	& Y \\

log $P_{orb}$& $D_{D}$ (1)		& -1.1 	&  1.8		& 7.9 	&  3.0		&  4.76		&  -0.07	&  0.00	& N \\
log $P_{orb}$& $D_{D}$ (2)		& 0.74  	&  0.11		& 1.72 	&  0.64		& 1.36		& 0.77 	& 0.59 	& Y \\

log $P_{orb}$& $D_{sq.w.}$ (1)	& -2.9  	&  1.2		& 18.5 	&  2.0		&  4.15		& -0.20 	& 0.04 	& Y \\
log $P_{orb}$& $D_{sq.w.}$ (2)	& 1.44  	&  0.42		& 4.8 	&  2.4		&  5.49		& 0.50 	& 0.25 	& Y \\

  \hline
\end{tabular}
\label{tab-corr}
\end{table*}


\bsp	
\label{lastpage}
\end{document}